\documentclass[11pt,a4paper]{article}
\usepackage[utf8]{inputenc}
\usepackage[T1]{fontenc}
\usepackage{amsmath}
\usepackage{amsfonts}
\usepackage{amssymb}
\usepackage{graphicx}
\usepackage{eurosym}
\usepackage{xcolor}
\usepackage{caption}
\usepackage{booktabs}
\usepackage{multirow}
\usepackage{siunitx}
\usepackage[left=2.50cm, right=2.50cm, top=2.50cm, bottom=2.50cm]{geometry}
\usepackage[onehalfspacing]{setspace}
\usepackage{placeins}
\usepackage{url}

\usepackage{apacite} 				
\usepackage{natbib}	
\usepackage{authblk}

\title{Analysing Opportunity Cost of Care Work using Mixed Effects Random Forests under Aggregated Auxiliary Data}
\author[*]{Patrick Krennmair}
\author[**]{Nora W{\"u}rz}
\author[**]{Timo Schmid}
\affil[*]{{\small Institute of Statistics and Econometrics, Freie Universität Berlin, Berlin, Germany}}
\affil[**]{{\small Institute of Statistics, Otto-Friedrich-Universität Bamberg, Bamberg, Germany}}
\affil[ ]{{\small Corresponding author: Nora W{\"u}rz, \texttt{nora.wuerz@uni-bamberg.de}}}
\date{}

\begin{document}
	
	\maketitle
	\onehalfspacing
	\normalsize
	\setlength{\parindent}{0em}
	
\begin{abstract}
	
Evidence-based policy-making requires reliable, spatially disaggregated indicators. The framework of mixed effects random forests leverages the advantages of random forests and hierarchical data in small area estimation. These methods require typically access to auxiliary information on population-level, which is a strong limitation for practitioners. In contrast, our proposed method - for point and uncertainty estimation - abstains from access to unit-level population data but adaptively incorporates aggregated auxiliary information through calibration-weights. We demonstrate its usage for estimating opportunity cost of care work for Germany from the Socio-Economic Panel and census aggregates. Simulation studies evaluate our proposed method.

\end{abstract}
	

\section{Introduction}\label{sec:1}
Evidence-based policy requires reliable socio-economic indicators. To analyse spatial aspects of inequality, we need precise information at a highly disaggregated level. When using survey data, an apparent trade-off is the inverse relation between high spatial resolution and correspondingly decreasing sample sizes. The estimation of indicators under these circumstances can be facilitated using an appropriate model-based methodology collectively referred to as Small Area Estimation (SAE) \citep{Rao_Molina2015, Tzavidis_etal2018}.
	
Models that exploit unit-level survey data to estimate area-level means are mainly regression-based linear mixed models, where the hierarchical structure of the observations is captured by random effects. A well-known model is the nested error regression model \citep{Battese_etal1988} - further labelled as BHF - which requires access to the survey and to area-level auxiliary information on the population. A versatile extension of the BHF model is the empirical best predictor by \citet{Molina_rao2010}. It differs from the BHF model as it enables the estimation of non-linear indicators, but requires access to population-level auxiliary data. The underlying linear mixed model of the BHF (and the empirical best predictor) relies on linearity and normality assumptions of the random effects and error terms, which can easily be violated in SAE applications. \citet{JiangRao2020} remind, that the quality of model-based SAE estimates is inevitably connected to the validity of model assumptions. Without accounting for these violations, point estimates are potentially biased and mean squared error (MSE) estimates are unreliable.
In SAE, several strategies evolved to prevent model-misspecification: A well-known example is the assurance of normality by transforming the dependent variable \citep{sugasawa2017transforming, Tzavidis_etal2018,  sugasawa2019adaptively, Rojas_etal2019}. Furthermore, the use of models under more flexible distributional assumptions is a fruitful approach \citep{DialloRao2018, Graf_etal2019}. Machine learning methods \citep{Hastie_etal2009, Varian2014} are especially beneficial because they avoid the parametric assumptions of  linear mixed models. \cite{Bilton_etal2017, Bilton_etal2020} estimate household poverty based on regression and classification trees. \cite{Krennmair_Schmid2021} recently introduce a framework for tree-based  machine learning methods (random forests) in SAE. From a different perspective, semi- or non-parametric approaches for the estimation of area-level means are investigated among others by \citet{Opsomeretal2008}, using penalized spline components within the linear mixed model setting.

All previously mentioned strategies against model-misspecification in SAE assume access to auxiliary information from unit-level population data. Due to data security reasons, the access to unit-level population data (for example census or register) is often strictly protected, constraining the methodological options of SAE practitioners. However, aggregated population-level auxiliary data (e.g., means) are often available at the spatial target level.
For point estimation of area-level means, \citet{Li_etal2019} propose the use of empirical likelihood based calibration weights and introduce a bias-corrected transformation approach using aggregated auxiliary data combined with the smearing approach of \citet{Duan1983}.   A further possibility to estimate area-level means and their uncertainty using log or log-shift transformation under limited access to population-level auxiliary data is presented in \citet{Wuerz_2022}. However, both approaches (\citet{Li_etal2019} and \citet{Wuerz_2022}) are still based on linear mixed models.

The starting point of this paper is the framework of \citet{Krennmair_Schmid2021}, which combines tree-based methods with perspectives from SAE. They propose a non-linear, data-driven, and semi-parametric method for estimating area-level means using mixed effects random forests (MERF). In general, a random forest $f$ (RF) \citep{Breiman2001} exhibits excellent predictive performance and implicitly addresses model-selection challenges. However, unlike linear models, unit-level auxiliary data $\mathbf{x}_{j}$ cannot simply be replaced by aggregated auxiliary information $\bar{\mathbf{x}}$ to form RF predictions, as $$f(\bar{\mathbf{x}})=f\left(\frac{1}{N} \sum_{j=1}^N \mathbf{x}_{j}\right) \neq \bar{f}(\mathbf{x}_{j})=\frac{1}{N} \sum_{j=1}^N f(\mathbf{x}_{j}).$$
Our methodological contributions address this critical limitation for practitioners and are twofold:
\begin{itemize}
	\item We combine the method proposed by \cite{Krennmair_Schmid2021} with calibration methods to enable the estimation of area-level means using the MERF (point estimation). Specifically, we introduce a strategy for adaptively incorporating aggregated auxiliary information through calibration-weights. While different calibration methods are feasible, we base the determination of the calibration-weights on the empirical likelihood approach \citep{Chen_Quin1993, QuinLawless1994, han_lawless2019}.
	\item We provide a (novel) integration of two existing bootstrap schemes, introduced by \citet{Chambers_Chandra2013} and \citet{Gonzalez_etal2008}, to estimate the MSE of the MERF (uncertainty estimation). The proposed bootstrap captures the dependence structure of the data and accounts for the uncertainty arising from the estimation of the MERF model in Equation (\ref{mod1}), while also enabling the simulation of bootstrap populations despite the use of aggregated auxiliary information.
\end{itemize}
These two contributions make the MERF applicable in situations where only aggregated population-level auxiliary data are available. To assess the strengths and weaknesses of our approach for point and uncertainty estimates, we compare it with existing SAE methods under scenarios with limited population-level auxiliary information in a model-based simulation study. 

From an applied perspective, we demonstrate our methodology using the $2021$ Socio-Economic Panel \citep{SOEP_datasource} combined with aggregated auxiliary information from the 2022 census to estimate the average individual opportunity cost of care work for $95$ regional planning regions (German: Raumordnungsregionen) in Germany. We refer to care work as unpaid working hours attributed to child- or elderly-care reported by the Socio-Economic Panel. Opportunity cost is an economic concept comprising the time allocation problem, where the time allocated for care work implicitly corresponds to time not providing paid work \citep{Buchanan1991}. Informally provided care work has no direct corresponding monetary value. Classical interpretations of labour supply in economics such as \cite{Becker1965}, imply that an individual’s hourly wage is an acceptable approximation of the unknown opportunity cost of time for working population.
Thus, we measure time cost by multiplying an individual’s care time by the opportunity cost of the person's time represented by the individual’s hourly wage from the Socio-Economic Panel. Unpaid care work mitigates public and private expenses on needed health services and infrastructure \citep{Chareles_eatl2005}. On the other hand, care-giving has a complex impact on the labour market \citep{Truskinovsky_Maestas2018, Stanfors_etal2019}, for instance by affecting workforce individuals through personal or social burdens \citep{Bauer_Sousa2015}. From a macro-perspective, several studies examine the economic value of care work for countries through the concept of opportunity cost \citep{Chari_etal2015, Ochalek_etal2018, mudrazija_etal2019} and provide empirical evidence for policy measures.
	
While spatial patterns of income inequality in Germany is of scientific interest \citep{frick2008regionalincome, kosfeld2008DisparitiesIncomeGermany, fuchs2010inequalityGermany}, to the best of our knowledge, no study on regional dispersion of opportunity cost of unpaid care work exists.
From a spatial perspective, \citet{OlivaMoreno2019} provide estimates on the economic value of time of informal care for two regions in Spain. We maintain that mapping opportunity cost of care work is particularly interesting for Germany as German Reunification and Federalism has resulted in powerful regional jurisdictions and different laws affecting care work. The visualization of opportunity cost highlights regional patterns, adding insights for planning and comparison of social-compensation policies.
	
The rest of the paper is structured as follows: Section \ref{sec:2.1} states a general mixed model that enables the use of tree-based models in SAE. Section \ref{sec:2.2} describes our area-level mean estimator based on MERFs under limited access to population-level auxiliary data. We scrutinize the use of empirical likelihood calibration weights and subsequently address methodological limitations in Section \ref{sec:2.3}. As a result, we propose a best practice strategy to ensure the proper usability of empirical likelihood calibration weights in the context of SAE. Section \ref{sec:3} introduces a non-parametric bootstrap-scheme to estimate the area-level MSE. In Section \ref{sec:4}, we evaluate the performance of our methods (point and MSE estimates) in model-based simulations under complex settings against existing competitors. The application (Section \ref{sec:5}) focuses on the estimation of average individual opportunity cost of care work for $95$ regions in Germany using the $2021$ Socio-Economic Panel. After the introduction of data sources and direct estimates in Section \ref{sec:5.1}, we address the gain in accuracy for the proposed method compared to direct and other SAE estimates under limited access to population-level auxiliary data. In Section \ref{sec:6}, we conclude and motivate further research.
	
\section{{Mixed Effects Random Forests}}\label{sec:2}
This section introduces a general mixed model enabling a simultaneous discussion of traditional linear mixed models in SAE such as the model of \citet{Battese_etal1988} as well as semi-parametric interpretations such as the model of \citet{Krennmair_Schmid2021} using MERFs. Section \ref{sec:2.2} provides details on our proposed methodology for MERFs under limited access to population-level auxiliary data and the determination of area-specific calibration weights based on empirical likelihood. We close the section with a discussion on limitations of empirical likelihood for SAE and state a best practice strategy ensuring the usability of our proposed point estimator in challenging empirical examples.
	
\subsection{Model and Estimation of Coefficients}\label{sec:2.1}
We assume a finite population $U$ of size $N$ consisting of $D$ separate domains $U_1, U_2,...,U_D$ with $N_1,N_2,...,N_D$ units, where index $i = 1,...,D$ indicates respective areas. The continuous target variable $y_{ij}$ for unit $j$ in area $i$ is available for every unit within the sample. Sample $s$ is drawn from $U$ and consists of $n$ units partitioned into sample sizes $n_1, n_2,..., n_D$ for all $D$ areas. We denote by $s_i$ the sub-sample from area $i$. The vector $\mathbf{x}_{ij} = (x_1, x_2,..., x_p)^\intercal$ includes $p$ explanatory variables and is available for every unit $j$ within the sample $s$. The relationship between $\mathbf{x}_{ij}$ and $y_{ij}$ is assumed to follow a general mixed effects regression model:
	
\begin{equation} \label{mod1}
	y_{ij} = f(\mathbf{x}_{ij}) + u_i + e_{ij} \quad \text{with}\quad u_i \sim N(0, \sigma^2_u) \quad \text{and} \quad e_{ij} \sim N(0,\sigma^2_e).
\end{equation}
	
Function $f(\mathbf{x}_{ij})$ models the conditional mean of $y_{ij}$ given $\mathbf{x}_{ij}$. The area-specific random effect $u_i$ and the unit-level error $e_{ij}$ are assumed to be independent. For instance, defining $f(\mathbf{x}_{ij}) = \mathbf{x}_{ij}^\intercal\mathbf{\beta}$ with $\mathbf{\beta} = (\beta_1,...,\beta_p)^\intercal$ and adding the intercept $\beta_0$ coincides with the well-known nested error regression model of \citet{Battese_etal1988} labelled as BHF.
An empirical best linear unbiased predictor for the area-level mean $\mu_i$ can be expressed as:
$$ \hat{\mu}^{\text{BHF}}_i = \hat{\beta}_0 + \bar{\mathbf{x}}_{\text{pop},i}^\intercal \hat{\mathbf{\beta}} + \hat{u}_i ,$$

where $\bar{\mathbf{x}}_{\text{pop},i} = \frac{1}{N_i} \sum_{j \in U_i} \mathbf{x}_{ij}$ denotes area-specific population means on $p$ covariates.
In a variety of real-world examples, it is very difficult to meet the model assumptions (e.g. linearity, normality assumption of the error terms), so that non-parametric approaches are an obvious choice \citep{JiangRao2020}. Tree-based machine learning methods such RFs \citep{Breiman2001} are data-driven procedures identifying predictive relations from data, including higher order interactions between covariates, without explicit model assumptions \citep{Hastie_etal2009, Varian2014}.  

Defining $f$ in Model (\ref{mod1}) to be a RF results in the a semi-parametric framework of MERF, which combines the advantages of RFs with the ability to model hierarchical structures of survey data using random effects. \citet{Krennmair_Schmid2021} estimate area-level means with RFs \citep{Breiman2001} and therefore introduce methodology to estimate the model-components $\hat{f}$, $\hat{u}_i$, $\hat{\sigma}_u^2$, and $\hat{\sigma}_e^2$ in the context of SAE.
For fitting Model (\ref{mod1}) (where $f$ is a RF) on survey data, the MERF algorithm subsequently estimates a) the forest function, assuming the random effects term to be correct and b) estimates the random effects part, assuming the Out-of-Bag-predictions from the forest to be correct. These predictions utilize the unused observations from the construction of each forest's sub-tree \citep{Breiman2001, Biau_Scornet2016}. The estimation of variance components $\hat{\sigma}^2_{e}$ and $\hat{\sigma}_u^2$ is obtained implicitly by taking the expectation of machine learning estimators given the data. For further methodological details, we refer to \citet{Krennmair_Schmid2021}. The resulting estimator for the area-level mean for MERFs is summarized as:

\begin{align}\label{MERForig}
\hat{\mu}_i^{\text{MERF}} = \bar{\hat{f}}_i(\mathbf{x}_{ij}) + \hat{u}_i, 
\quad \quad \text{where} \quad &\bar{\hat{f}}_i(\mathbf{x}_{ij}) = \frac{1}{N_i} \sum_{j \in U_i} \hat{f}(\mathbf{x}_{ij}).
\end{align}

\subsection{MERFs under Aggregated Auxiliary Information}\label{sec:2.2}
Estimates for the area-level mean $\mu_i$ using MERFs from Equation (\ref{MERForig}) require unit-level auxiliary population data as input for $f$.
In contrast to the linear BHF model, aggregated area-specific auxiliary information ($\bar{\mathbf{x}}_{\text{pop},i}$) cannot directly be used for non-linear or non-parametric procedures such as RFs, as in general $f(\bar{\mathbf{x}}_{\text{pop},i}) \neq \bar{f}_i(\mathbf{x}_{ij})$.
Although the access to unit-level auxiliary population data for the covariates imposes a limitation for practitioners, not many strategies to prevent model-misspecification in SAE consider simultaneously limited access to population-level auxiliary data. We propose a solution overcoming this issue by calibrating model-based estimates from MERFs in Equation (\ref{MERForig}) with weights that are based only on aggregated population-level auxiliary information (means). The general idea originates from the bias-corrected transformed nested error regression estimator using aggregated auxiliary data (\textit{TNER2}) by \cite{Li_etal2019}. We build on their idea of using calibration weights for SAE based on empirical likelihood and transfer it to MERFs. As a result, our proposed method offers benefits of RFs such as good predictive performance and implicit model-selection, while simultaneously working in cases of limited access to population-level auxiliary data. In short, our estimator for the area-level mean can be written as:

\begin{equation}\label{MERFagg}
	\hat{\mu}_i^{\text{MERFagg}}
	= \sum_{j = 1}^{n_i} \hat{w}_{ij} \left[ \hat{f}(\mathbf{x}_{ij}) + \hat{u}_i \right].
\end{equation}

Note that optimal estimates for required model-components $\hat{f}$ and $\hat{u}_i$ are obtained similar to Equation (\ref{MERForig}) from survey data using the MERF algorithm as described by \citet{Krennmair_Schmid2021}. We incorporate aggregated population-level auxiliary information from a census through the calibration weights $w_{ij}$, which balance unit-level predictions to achieve consistency with the area-wise aggregates (means) from census data.
Following \citet{owen1990empirical} and \citet{QuinLawless1994} the technical conditions for $w_{ij}$ are to maximize the profile empirical likelihood function $ \prod_{j = 1}^{n_i} w_{ij}$ under the following three constraints:
\begin{itemize}
 \item$\sum_{j = 1} ^{n_i} w_{ij}(\mathbf{x}_{ij} - \bar{\mathbf{x}}_{\text{pop},i}) = \mathbf{0}$,
 here the population-level covariate means ($ \bar{\mathbf{x}}_{\text{pop},i}$) are used for calibration. Thus, a system of $p$ equations must be solved that consist of the sum over the product of weights and distances between survey data and the population-level mean;
 \item $ w_{ij} \geq 0$, ensuring the non-negativity of weights;
 \item $\sum_{j = 1}^{n_i} w_{ij} = 1$, to normalize weights.
\end{itemize}
Optimal weights $\hat{w}_{ij}$, maximizing the profile empirical likelihood under the given constraints, are found by the Lagrange multiplier method:
\begin{align}
	\label{equ:LiLiu_weights}
	\hat{w}_{ij} = \frac{1}{n_i} \frac{1}{1+\hat{\lambda}_i^\intercal(\mathbf{x}_{ij} - \bar{\mathbf{x}}_{\text{pop},i}) }, \\
	\text{where}\enspace \hat{\mathbf{\lambda}}_i \enspace \text{solves} \enspace \sum_{j = 1}^{n_i} \frac{\mathbf{x}_{ij} - \bar{\mathbf{x}}_{\text{pop},i}}{1+ \hat{\mathbf{\lambda}}_i^\intercal (\mathbf{x}_{ij} - \bar{\mathbf{x}}_{\text{pop},i})} = \mathbf{0}. \notag
\end{align}
An important competitor to the $MERFagg$ in Equation (\ref{MERFagg}) for estimating area-level means is a synthetic estimator based on a classical RF. This estimator is constructed by fitting a RF to the sample data without accounting for area-specific random effects $u_i$. It can be expressed as:
\begin{equation}\label{RFagg}
	\hat{\mu}_i^{\text{RFagg}}
	= \sum_{j = 1}^{n_i} \hat{w}_{ij} \hat{f}(\mathbf{x}_{ij}).
\end{equation}
We use $RFagg$ in our simulation studies to investigate whether including area-specific random effects in MERFs results in efficiency gains compared to a purely synthetic RF-based estimator.

\subsection{Limitation of Empirical Likelihood and a Best Practice Advice for SAE}\label{sec:2.3}

The existence of a solution to the maximization problem for the calibration weights $\hat{w}_{ij}$ is not necessarily guaranteed for applications in SAE. A necessary and sufficient condition ensuring the existence of a solution for $\hat{\mathbf{\lambda}}_i$ is that the convex hull of the constraint matrix $\mathbf{x}_{ij} - \bar{\mathbf{x}}_{\text{pop},i}$ contains the origin as an interior point. Especially for small sample sizes $n_i$ this condition requires scrutiny \citep{Emerson_Owen2009}. If the sample means of $\mathbf{x}_{ij}$ for area $i$ strongly differ from $\bar{\mathbf{x}}_{\text{pop},i}$, for instance, due to a strong imbalance of the unit-level sample values $\mathbf{x}_{ij}$ around the area-specific mean from population data $\bar{\mathbf{x}}_{\text{pop},i}$, no optimal solution for $\hat{\mathbf{\lambda}}_i$ and subsequently $\hat{w}_{ij}$ cannot be obtained.

We propose a stepwise approach to ensure a solution to Equation (\ref{equ:LiLiu_weights}) for each area $i$. This approach can be interpreted as a best-practice strategy for incorporating the maximum auxiliary covariate information through calibration weights in Equation (\ref{equ:LiLiu_weights}) when estimating area-level means with MERFs. Our strategy for addressing potential failures in weight calculation is summarized in the following algorithmic framework.

\par\noindent\rule{\textwidth}{0.4pt}
\begin{enumerate}
	\item Use MERF to obtain estimates $\hat{f}$, $\hat{u}_i$, $\hat{\sigma}_u^2$, and $\hat{\sigma}_e^2$ from available unit-level survey data and estimate the indicator $\hat{\mu}_i^{\text{MERFagg}}$ (\ref{MERFagg}) including weights $\hat{w}_{ij}$ following Equation (\ref{equ:LiLiu_weights}).
	\item If the calculation of weights fails due to infeasibility of constraints in the optimization problem for area $i$:
	\begin{enumerate}
	\item Check the feasibility of constraints used in the optimization and remove perfectly co-linear columns in $(\mathbf{x}_{ij} - \bar{\mathbf{x}}_{\text{pop},i})_{j = 1, ..., n_i}$. Retry the optimization in Equation (\ref{equ:LiLiu_weights}).
	\item If the calculation of weights fails again, optionally enhance the domain-specific sample size of area $i$ by sampling randomly with replacement from the most ``similar'' domain according to the minimal row-wise Euclidean distance between area-specific aggregated auxiliary  information $\bar{\mathbf{x}}_{\text{pop},i}$. Retry the calculation of weights $\hat{w}_{ij}$.
	\item If it fails again, reduce the number of covariates used for the calculation of weights for area $i$. Starting with the least influential covariate based on variable importance from $\hat{f}$, reduce the number of covariates in each step and retry the calculation of weights after each step.
	\item If the calculation of weights was not possible in step (c), set $\hat{w}_{ij}$ to $1/n_i$. These weights are non-informative for incorporating auxiliary information, however, the model-based estimates $\hat{f}(\mathbf{x}_{ij}) + \hat{u}_i$ still comprise information from other in-sample areas.
	\end{enumerate}
	\item Calculate the indicator for the $i$-th area as proposed by Equation (\ref{MERFagg}).
\end{enumerate}
\par\noindent\rule{\textwidth}{0.4pt}

Please note that a more detailed discussion of the maximization problem, along with additional details about the algorithmic framework, is provided in Section 2 of the online supplementary materials. The general performance of the best-practice strategy is illustrated by the results of the model-based simulation in Section \ref{sec:4}. Additionally, the proposed best-practice strategy is applied in the application discussed in Section \ref{sec:5}.

An alternative strategy to avoid potential issues with maximizing the profile empirical likelihood function under the given constraints is to consider alternative calibration methods. Instead of reducing the number of covariates used for weight calculation in Step 2 (c) of the algorithm, ridge calibration \citep{rao1997ridge, bocci2008another, montanari2009multiple} modifies the optimization problem by introducing a penalty term. This approach prevents extreme weights and improves numerical stability by relaxing the linear constraints on the covariates. Another method to stabilize calibration weights at the area-level (and to avoid numerical instabilities) involves calculating the calibration weights at a higher (hierarchical) level, under the assumption that the covariates at the area-level are homogeneous within the respective higher levels in which they are nested. Both alternatives are assessed and compared to the proposed best-practice strategy in the application in Section \ref{sec:5}.

\section{Uncertainty Estimation}\label{sec:3}
The area-wise MSE is a conventional measure to assess the uncertainty of provided point estimates. While the quantification of uncertainty is essential for determining the quality of area-level estimates, its calculation remains a challenging task.
An analytical approximation, as for the BHF model \citep{Prasad_Rao1990, Datta_Lahiri2000}, is not trivial due to the non-linearity of the RF. Thus, the estimation of uncertainty by elaborate bootstrap-schemes is an established alternative \citep{Hall_Maiti2006, Gonzalez_etal2008, Chambers_Chandra2013}.

In this paper, we propose a non-parametric bootstrap for finite populations estimating the MSE of the introduced area-level estimator under limited auxiliary information defined by Equation (\ref{MERFagg}). Essentially, we aim to find a solution to two problems simultaneously: Firstly, we need to capture the dependence-structure of the data and uncertainty introduced by the estimation of Model (\ref{mod1}). Secondly, we face problems in simulating a full bootstrap population in the presence of aggregated auxiliary information from census data.

Our proposed solution to this dual problem is the combination of two existing bootstrap schemes introduced by \citet{Chambers_Chandra2013} and \citet{Gonzalez_etal2008}. \citet{Chambers_Chandra2013} address the problem of non-parametric generation of random components. One key-advantage is its leniency to potential specification errors of the covariance structure, as the extraction of the empirical residuals only depends on the correct specification of the mean behavior function $f$ of the model. To solve the problem of missing unit-level population auxiliary data, we base the general procedure on the parametric bootstrap for finite populations introduced by \citet{Gonzalez_etal2008}. This allows us to find (pseudo-)true values by generating only error components instead of simulating full bootstrap populations at unit-level. An important step concerning the handling and resampling of empirical error components is centering and scaling them by a bias-adjusted residual variance $\hat{\sigma}^2_{bc,e}$ proposed by \citet{Mendez_Lohr2011}. In short, the estimator of the residual variance under the MERF from Equation (\ref{MERForig}), $\hat{\sigma}^2_{e}$ is positively biased, as it includes excess uncertainty concerning the estimation of function $\hat{f}$. Further methodological details on the modification of the approach by \citet{Chambers_Chandra2013} for MERFs for area-level means under unit-level models are found in \citet{Krennmair_Schmid2021}. The steps of the proposed bootstrap are as follows:

\par\noindent\rule{\textwidth}{0.4pt}
\begin{enumerate}
	\item Use estimates $\hat{f}$, $\hat{\sigma}_{e}^2$, $\hat{\sigma}_{u}^2$, and respective weights $\hat{w}_{ij}$ from the application of the proposed method as summarized in Equation (\ref{MERFagg}) on survey data with metric target variable $y_{ij}$.
	\item Calculate marginal residuals $\hat{r}_{ij} = y_{ij} -\hat{f}(\mathbf{x}_{ij})$ and use them to compute level-2 residuals for each area by $\bar{r}_{i} = \frac{1}{n_i} \sum_{j=1}^{n_i} {\hat{r}_{ij}}$ for $i=1,...D$.
	\item To replicate the hierarchical structure we use the marginal residuals and obtain level-1 residuals by $r_{ij} = \hat{r}_{ij} - \bar{r}_i$. Level-1 residuals $r_{ij}$ are scaled to the bias-corrected variance $\hat{\sigma}_{\text{bc},e}^2$ \citep{Mendez_Lohr2011} and centered, denoted by $r^{c}_{ij}$. Level-2 residuals $\bar{r}_i$ are also scaled to the estimated variance $\hat{\sigma}_{u}^2$ and centered, denoted by $\bar{r}^{c}_i$.
	\item For $b=1,...,B$:
	\begin{enumerate}
		\item Three random components ($r_{ij}^{*(b)}$,$\bar{e}_i^{*(b)}$,$u^{*(b)}_i$) are sampled using random sampling with replacement (srswr) for each area $i$ for every replication $b$:
		\begin{itemize}
			\item sample $n_i$ level-1 residuals for each area $i$ from the scaled and centered empirical level-1 residuals $r^c_{ij}$  $$r_{ij}^{*(b)}=srswr(r^c_{ij},n_i)$$
			\item sample one value for each area $i$ to consider the averaged deviation produced by out-of-sample parts on level-1
			$$\bar{e}_i^{*(b)} = srswr(r^c_{ij}\frac{\hat{\sigma}_{\text{bc},e}}{\sqrt{N_i-n_i}},1)$$
			\item sample one level-2 residual for each area $i$ from the scaled and centered empirical level-2 residuals $\bar{r}^c$
			$$u^{*(b)}_i=srswr(\bar{r}^c,1)$$
		\end{itemize}
		\item Compute (pseudo-)true values for the population based on the fixed effects from area-wise mean estimates $\hat{\mu}_i^{\text{MERFagg}}$, as:
		\begin{eqnarray}
			\nonumber \bar{y}_i^{(b)} = \sum_{j=1}^{n_i}\hat{w}_{ij}\hat{f}(\mathbf{x}_{ij}) + u_i^{*(b)} + \bar{E}^{(b)}_i, \enspace \enspace\enspace \text{where} \enspace\enspace\enspace
			\bar{E}^{(b)}_i = \frac{n_i}{N_i} \bar{r}_{ij}^{*(b)} + \frac{N_i -n_i}{N_i} \bar{e}_i^{*(b)}.
		\end{eqnarray}
		The random component $\bar{E}^{(b)}_i$ adds the mean deviation for each area $i$ coming from the level-1 residuals and therefore $\bar{E}^{(b)}_i$ is a weighted sum of averaged sampled empirical level-1 residuals ($\bar{r}_{ij}^{*(b)}$) and out-of-sample averaged deviation ($\bar{e}_i^{*(b)}$).
		\item Use the known sample covariates $\mathbf{x}_{ij}$ to generate the bootstrap sample response values in the following way:
		$$y_{ij}^{(b)} = \hat{f}^{\text{OOB}}(\mathbf{x}_{ij}) + u^{*(b)}_i + r_{ij}^{*(b)}.$$ We use Out-of-Bag-predictions $\hat{f}^{\text{OOB}}$ from $\hat{f}$ to capture variations of covariates $\mathbf{x}_{ij}$ through predictions from unused observations within each tree in the fitting process that vary throughout the bootstrap replications.
		\item Estimate $\hat{\mu}_i^{\text{MERFagg}(b)}$ with the proposed method from Equation (\ref{MERFagg}) on bootstrap sample values $y_{ij}^{(b)}$.
	\end{enumerate}
		\item Finally, calculate the estimated MSE for the area-level mean for areas $i = 1,...,D$
	$$\widehat{\text{MSE}}\left(\hat{\mu}_i^{\text{MERFagg}}\right) = \frac{1}{B} \sum_{b = 1}^B \left[\left(\hat{\mu}_i^{\text{MERFagg}(b)} - \bar{y}_i^{(b)}\right)^2\right].$$
\end{enumerate}
\par\noindent\rule{\textwidth}{0.4pt}\\

\section{Model-Based Simulation}\label{sec:4}
The model-based simulation allows for a controlled empirical assessment of our methods for point and uncertainty estimates. Overall, we aim to show, that the proposed methodology from Section \ref{sec:2} and Section \ref{sec:3} performs as well as traditional SAE methods and has advantages in terms of robustness against model-failure. In particular, we study the performance of MERFs under limited auxiliary data access (\textit{MERFagg}, (\ref{MERFagg})) to a synthetic estimator based on a classical RF under limited auxiliary data access without accounting for area-specific random effects (\textit{RFagg}, (\ref{RFagg})) as well as the MERF assuming access to unit-level auxiliary data (\textit{MERFind}, (\ref{MERForig})) by \citet{Krennmair_Schmid2021}. The inclusion of \textit{RFagg} allows for a comparison to determine whether incorporating area-specific random effects in MERFs leads to efficiency gains compared to a purely synthetic RF-based estimator. Additionally, the inclusion of \textit{MERFind} facilitates a direct comparison of the impact of using aggregated population-level auxiliary data (\textit{MERFagg}) versus unit-level population-level auxiliary data (\textit{MERFind}). These estimators are compared to four alternatives that use only aggregated population-level auxiliary data: a) the \textit{BHF} estimator by \cite{Battese_etal1988}, b) the \textit{BHFls} estimator, which incorporates an adaptive log-shift transformation, as proposed by \cite{Wuerz_2022}, c) the \textit{TNER2} estimator introduced by \cite{Li_etal2019}, and d) the area-level \textit{FH} estimator by \cite{Fay_Heriot1979}. The \textit{BHF} model serves as a well-established baseline for estimating area-level means under limited population-level auxiliary data. The \textit{BHFls} and \textit{TNER2} estimators aim to provide alternatives to the \textit{BHF}, incorporating (adaptive) transformations under limited auxiliary data access. Differences in performance between these estimators and the \textit{MERFagg} highlight the potential advantages of semi-parametric and non-linear modelling in the given data scenarios. Please note that Table 1 in the online supplementary materials contains a list of abbreviations for reference.

We consider four scenarios denoted as \textit{Normal}, \textit{Pareto}, \textit{Interaction}, and \textit{Logscale} and repeat each scenario independently $M=500$ times. All four scenarios assume a finite population $U$ of size $N=50000$ with $D=50$ disjunct areas $U_1,...,U_D$ of equal size $N_i = 1000$. We generate samples under stratified random sampling, utilizing the $50$ small areas as stratas, resulting in a sample size of $n = \sum_{i=1}^{D} n_i = 1229$. The area-specific sample sizes range from $5$ to $50$ sampled units with a median of $21$ and a mean of $25$. The sample sizes are comparable to area-level sample sizes in the application in Section \ref{sec:5} and can thus be considered to be realistic.

The choice of the simulation scenarios is motivated by our aim to evaluate the performance of the competing methods for economic and social inequality data. This includes skewed data, deviations from normality of error terms, or the presence of unknown non-linear interactions between covariates, that might trigger model-misspecifications in traditional SAE approaches based on  linear mixed models.
The data generating processes for the used scenarios are provided in Table \ref{tab:MB1}.
Scenario \textit{Normal} provides a baseline under a  linear mixed model with normally distributed random effects and unit-level errors. As the model assumptions for  linear mixed models are fully met, we aim to show that the \textit{MERFagg} performs similarly well compared to linear competitors.
Scenario \textit{Pareto} is based on the same linear additive structure as scenario \textit{Normal}, but has Pareto distributed unit-level errors. This leads to a skewed target variable, comparable to empirical cases of monetary data. The data generating process of scenario \textit{Interaction} likewise results in a skewed target variable $y_{ij}$, although it shares its structure of random components with \textit{Normal}. The \textit{Interaction} scenario portrays advantages of semi-parametric and non-linear modelling methods protecting against model-failure arising from models with unknown interactions.
Scenario \textit{Logscale} introduces an additional example resulting in a skewed target variable. Log-normally distributed variables mimic realistic income scenarios and constitute a showcase for SAE transformation approaches. We want to show the ability of MERFs and particularly of \textit{MERFagg} to handle such scenarios as well by identifying the non-linear relation introduced trough the transformation on the linear additive terms.

\begin{table}[tb]
	\centering
	\captionsetup{justification=centering,margin=1.5cm}
	\caption{Model-based simulation scenarios}
	\resizebox{\textwidth}{!}{\begin{tabular}{rlccccccc}
			\toprule \toprule
			{Scenario} & {Model} & {$x1$} & {$x2$} & {$\mu_i$} & {$v$} & {$e$} \\ \midrule
			Normal  & $ y = 5000-500x_1-500x_2+v+e$ & $N(\mu_i,3^2)$ & $N(\mu_i,3^2)$ & $unif(-1,1)$  & $N(0,500^2)$ & $N(0,1000^2)$  \\
			Pareto  & $ y = 5000-500x_1-500x_2+v+e $  & $N(\mu_i,3^2)$ & $N(\mu_i,3^2)$  & $unif(-1,1)$ & $N(0,500^2)$ &$Par(3,800^2)$ \\
			Interaction  & $ y = 1000 + 100x_1x_2 +75x_2 + v + e$  & $N(\mu_i,2^2)$ & $N(\mu_i,1)$  & $unif(-7,7)$ & $N(0,500^2)$ &  $N(0,1000^2)$\\
			Logscale  & $ y = \exp(7.5 - 0.25x_1 - 0.25 x_2 + v + e)$  & $N(\mu_i,1)$ & $N(\mu_i,1)$  & $unif(-3,3)$ & $N(0,0.15^2)$  & $N(0,0.25^2)$ \\ \bottomrule
	\end{tabular}}
	\label{tab:MB1}
\end{table}

We evaluate point estimates for the area-level mean over $M$ replications by the empirical root MSE (RMSE) and the relative bias (RB). As quality-criteria for the evaluation of the MSE estimates, we choose the relative bias of RMSE (RB-RMSE):
\begin{align}\text{RMSE}_i = \sqrt{\frac{1}{M} \sum_{m=1}^{M}(\hat{\mu}^{(m)}_i -\mu^{(m)}_i)^2}\\
\text{RB}_i = \frac{1}{M} \sum_{m=1}^{M} \left(\frac{\hat{\mu}^{(m)}_i - \mu^{(m)}_i}{\mu^{(m)}_i}\right),\\
\text{RB-RMSE}_i =\frac{\sqrt{\frac{1}{M} \sum_{m=1}^{M} MSE^{(m)}_{\text{est},i}} - RMSE_i}{RMSE_i},\end{align}
where $\hat{\mu}^{(m)}_i$ is the estimated mean in area $i$ based on any of the methods mentioned above and $\mu^{(m)}_i$ defines the true mean for area $i$ in replication $m$. $MSE_{\text{est},i}^{(m)}$ is estimated by the proposed bootstrap from Section \ref{sec:3}.

For the computational realization of the model-based simulation, we use \textsf{R} \citep{R_language}. The \textit{BHFls} and \textit{BHF} estimates are realized from the \textsf{saeTrafo}-package \citep{saeTrafoPackage}. For the estimates of the \textit{TNER2}, we used code provided by \cite{Li_etal2019}. For estimates based on the MERF approach, we use the package \textsf{SAEforest} \citep{SAEforestPackage} and for the \textit{RFagg} we use the \textsf{ranger}-package \citep{WrightZiegler2017}. For RFs, we set the number of split-candidates to $1$, keeping the default of $500$ trees for each forest. The \textit{FH} estimates are based on the package \textsf{emdi} \citep{Harmening2023}.

\subsection{Performance of Point Estimators of the Small Area Means}

\begin{figure}[tb]
	\centering
	\captionsetup{justification=centering,margin=1.5cm}
	\includegraphics[width=0.9\linewidth]{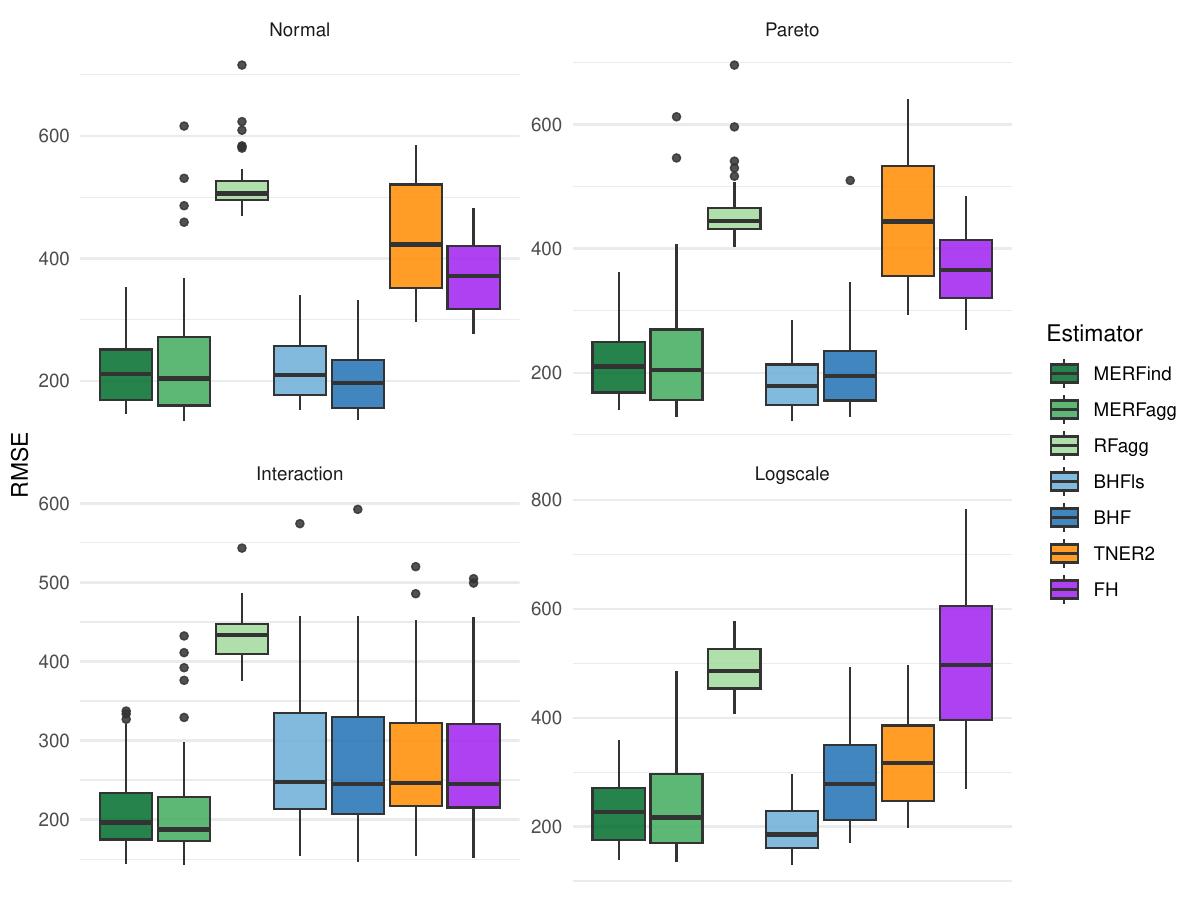}
	\caption{ Empirical RMSE comparison of point estimates for area-level averages under four scenarios.}
	\label{fig:MBpoint}
\end{figure}

Figure \ref{fig:MBpoint} reports the RMSE of each point estimation method under the four scenarios. Additional results are provided in Section 3 of the online supplementary materials, where Figure 1 shows the RB, and Table 2 summarizes the mean and median values across areas for both measures.

In the \textit{Normal} scenario, the \textit{BHF} performs best as it replicates the data generating process. The inclusion of the \textit{BHFls} estimator is for demonstrating how an adaptive log-shift transformation adapts to the data generation mechanism. The \textit{MERFind} and the \textit{MERFagg} perform on a comparable level, which shows that the proposed calibration approach to incorporate aggregated auxiliary information works. Although the MERF-based estimators are less efficient than the $BHF$ estimator under normality assumptions, they maintain a competitive performance overall. Similar patterns appear in the \textit{Pareto} scenario, with both MERF-based methods (\textit{MERFind} and \textit{MERFagg}) performing comparably to the \textit{Normal} setting, confirming robustness to skewed data and non-normal errors. Since \textit{MERFagg} behaves comparably, the robustness also holds for the calculation of calibration weights. In the \textit{Interaction} scenario, the proposed \textit{MERFagg} estimator outperforms traditional SAE approaches that rely solely on aggregated population-level auxiliary data. This indicates that linear mixed model-based methods struggle to adequately capture the underlying predictive relationships between covariates, whereas the MERFs successfully identify the non-linear terms. The importance of incorporating random effects is further emphasized by the poor performance of \textit{RFagg}, which fares even worse than its linear mixed model-based competitors. In the \textit{Logscale} scenario, both MERF-based methods (\textit{MERFind} and \textit{MERFagg}) work only slightly worse as the \textit{BHFls}, which can replicate the data generating process due to the inclusion of the adaptive log-shift transformation.

Overall, the MERF-based methods (\textit{MERFind} and \textit{MERFagg}) perform comparably well to linear mixed models in (linear) scenarios (\textit{Normal} and \textit{Pareto}), and outperforms in situations with unknown non-linear relations between covariates. The \textit{MERFagg} estimator outperforms the \textit{RFagg} estimator, demonstrating the advantage of incorporating random effects when unexplained area-level heterogeneity exists. The similarity of \textit{MERFind} and \textit{MERFagg} emphasizes that the influence of unit-level versus aggregated auxiliary information appears to be marginal in all of our four scenarios.

\subsection{Performance of the Bootstrap MSE Estimator}

We scrutinize the performance of our proposed MSE estimator on the four scenarios, examining whether the proposed procedure for uncertainty estimates performs equally well in terms of robustness against model-misspecification and in cases of limited access to population-level auxiliary information. 

\begin{table}[!t]
	\centering
	\captionsetup{justification=centering,margin=1.5cm}
	\caption{Performance of MSE estimator in model-based simulation: mean and median of RB-RMSE over areas.}
	\resizebox{\textwidth}{!}{\begin{tabular}{@{\extracolsep{5pt}} lcccccccc}
			\\[-1.8ex]\hline
			\hline \\[-1.8ex]
			&\multicolumn{2}{c}{\textit{Normal}} &\multicolumn{2}{c}{\textit{Pareto}}&\multicolumn{2}{c}{\textit{Interaction}}&\multicolumn{2}{c}{\textit{Logscale}} \\
			\hline \\[-1.8ex]
			& Median & Mean & Median & Mean & Median & Mean & Median & Mean \\
			\hline \\[-1.8ex]
			RB-RMSE & $0.0525$ & $0.0591$ & $0.0596$ & $0.0643$ & $0.0192$ & $0.0205$ & $-0.0117$ & $0.0054$ \\
			\hline \\[-1.8ex]
	\end{tabular}}
	\label{tab:MBmse}
\end{table}
For each scenario and each simulation round, we choose $B = 200$ bootstrap replications. From the comparison of RB-RMSE among the four scenarios provided in Table \ref{tab:MBmse}, we infer, that the proposed non-parametric bootstrap-procedure effectively handles all four scenarios. This is demonstrated by relatively low mean values of positive RB-RMSE over the $50$ areas after $M$ replications. From an applied perspective, we prefer over- to underestimation for the MSE as it serves as an upper bound.
The difference in RB-RMSE between \textit{Normal} and \textit{Pareto} is marginal, indicating that the non-parametric bootstrap effectively handles non-Gaussian error terms.

Figure 2 in the online supplementary materials assesses the area-wise tracking properties of our non-parametric bootstrap estimator. We conclude in all four scenarios, that our MSE estimates strongly correspond to the empirical RMSE. The overestimation in Table \ref{tab:MBmse} is mainly driven by overestimation in areas with low sample sizes. Thus, our non-parametric bootstrap provides conservative uncertainty estimates in areas with low sample sizes. Apart from this characteristic, we observe no further systematic differences between the estimated and empirical RMSE estimates regarding their performance throughout our model-based simulation.

\FloatBarrier
\section{Application}\label{sec:5}

This section starts with a description of the data sources and an outline of our empirical analysis. We describe the survey data (Socio-Economic Panel) and discuss primary \textit{direct} estimates of spatial differences in the average individual opportunity cost of care work for German regions. Moreover, we propose the use of model-based SAE, which incorporates auxiliary variables from the $2022$ census in Germany. By demonstrating our proposed method of MERFs with aggregated auxiliary information for point and uncertainty estimates, we highlight its advantages over existing model-based SAE methods. Finally, we discuss our empirical findings regarding the cost of care work in Germany. 

\subsection{Data Sources and Direct Estimates of Spatial Opportunity Cost of Care Work}\label{sec:5.1}
The Socio-Economic Panel was established in $1984$ by the German Institute of Economic Research (German: Deutsches Institut für Wirtschaftsforschung) and has become a vital survey for Germany, providing multidisciplinary social information on private households \citep{goebel2019german}. For our primary calculation of the opportunity cost of care work, we require information on individual income as well as hours worked in both paid employment and unpaid care work. This detailed information is only available in the Socio-Economic Panel, unlike the German Microcensus \citep{Bundesamt_2015}, where income is only available as an interval censored variable.

\begin{figure}[tb]
	\centering
	\captionsetup{justification=centering,margin=1.5cm}
	\includegraphics[width = \linewidth]{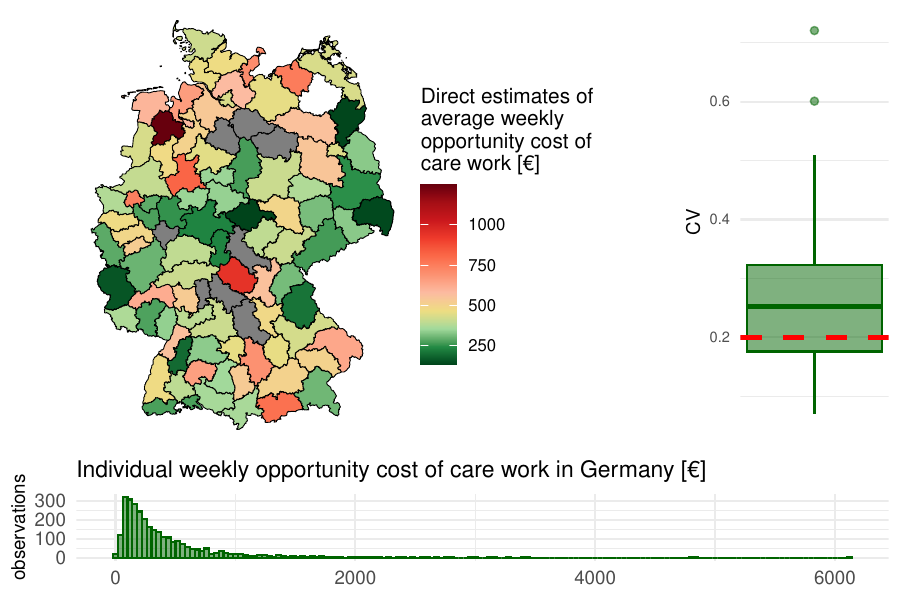}
	\caption{Overview of \textit{direct} estimates (areas with sample size under 10 are greyed out), corresponding CVs and the distribution of opportunity cost of care work in Germany.}
	\label{fig:Overviewdirect}
\end{figure}

We construct the target variable of individual weekly opportunity cost of care work using data from the Socio-Economic Panel in $2021$ \citep{SOEP_datasource}. We selected the year $2021$ because it is the  latest available wave of the Socio-Economic Panel. The sampling design follows a multi-stage stratified sampling procedure: stratification is first carried out by federal states, governmental regions, and municipalities. Subsequently, addresses are sampled using the random walk methodology within each primary sampling unit \citep{siegers2022soep}. Our analysis focuses on the working age population, defined as individuals aged between $15$ to $64$, in accordance with international standards \citep{OECD_2020}. Specifically, we calculate the individual opportunity cost in Euros per week for $2021$ as follows: first, the opportunity cost is computed as the hourly wage, derived by dividing mean gross individual income by hours of paid work. Then, the weekly hours of unpaid work due to child- or elderly-care are multiplied by the hourly opportunity cost. The resulting metric target variable $y_{ij}$ is highly skewed, ranging from \EUR$0$ to \EUR$6111.1$ (mean: \EUR$403.0$; median: \EUR$258.6$). A histogram is provided in Figure \ref{fig:Overviewdirect}.

In total, we have $2980$ individuals in the sample. Our major interest is to create a fine spatial resolution map to identify patterns in the opportunity cost of care work across Germany. We analyse $95$ regions in Germany (one regions could not be matched due to the district reform in Mecklenburg-Vorpommern), resulting in area-specific sample sizes ranging from $3$ to $145$, with a mean of $31.37$ and a median of $24$. First results of the \textit{direct} estimates are presented in the map in Figure \ref{fig:Overviewdirect}. Estimates of the mean weekly opportunity cost of individual care work range from \EUR$136.30$ (Uckermark-Barnim) to \EUR$1249.9$ (Oldenburg). In general, we observe no significant differences between former East and West Germany. Additionally, levels of opportunity cost are higher in metropolitan areas surrounding cities compared to the cities themself and rural areas.

However, the small sample sizes lead to unreliable direct estimates with high variances. Furthermore, we are not allowed to report \textit{direct} estimates for regions with a sample size below $10$ due to confidentiality agreements with the data provider. This applies to $8$ regions. To obtain variances and subsequently determine the coefficients of variation (CV) for the \textit{direct} estimates, we use the calibrated bootstrap by \citet{alfons2012estimation}, implemented in the \textsf{R}-package \textsf{emdi} by \citet{Kreutzmann_etal2019}. \cite{eurostatCV} postulates that estimates with a CV of less than $20\%$ can be considered as reliable. As reported by Figure \ref{fig:Overviewdirect}, more than half of the regions ($56$ out-of the remaining $87$) exceed this threshold.

The \textit{direct} estimation results suffer from differences in quality due to low area-level sample sizes and high variability. Model-based SAE methods help to improve the accuracy of these estimates. Since Socio-Economic Panel auxiliary variables are measured in the same way as in the German census \citep{Bundesamt_2015}, census covariate data can be used as auxiliary information in SAE models. However, the German census provides information only at aggregated regional levels. Overall, we have $19$ covariates on personal and socio-economic backgrounds within our sample, for which we also received corresponding means calculated by the German Statistical Office from the $2022$ German census. Details on available covariates and their variable importance are provided in Table 3 in the online supplementary materials.

\subsection{Model-Based Estimates}\label{sec:5.2}
This section illustrates the application of our proposed method for MERFs with aggregated auxiliary information to estimate area-level means. We map the estimated weekly mean opportunity cost of unpaid care work for $95$ regions in Germany for the year $2022$. Furthermore, we assess the quality of our estimates by providing CVs derived from our proposed non-parametric MSE-bootstrap procedure discussed in Section \ref{sec:3}. We then compare our results to the previously discussed \textit{direct} estimator, the $BHF$ estimator, the $BHFls$ estimator with an adaptive log-shift transformation, and the area-level \textit{FH} estimator. A full comparison with the \textit{TNER2} estimates is not possible because \citet{Li_etal2019} do not provide uncertainty estimators necessary for a qualitative comparison in terms of CVs.

As shown in Figure \ref{fig:Overviewdirect}, our target variable of individual opportunity cost is highly skewed, which increases the risk of model-misspecification in traditional linear mixed models (such as the BHF). In this context, the BHF with an adaptive log-shift transformation may help to protect against model-misspecification by applying the transformation. An alternative approach is our proposed procedure, which demonstrates robustness against model-failure caused by outliers or complex data structures. In addition to modeling separate regions as random intercepts, the proposed \textit{MERFagg} approach is purely data-driven: we train a predictive model on the survey data and incorporate as much auxiliary information as possible for determining area-specific calibration weights, $\hat{w}_{ij}$ (see Section \ref{sec:2.3}), based on the variable importance obtained from the fitted RF object $\hat{f}$. We set the tuning parameter of the RF to $500$ sub-trees. For the non-parametric MSE-bootstrap procedure, we use $B=200$.

\begin{figure}[tb]
	\centering
	\captionsetup{justification=centering,margin=1.5cm}
	\includegraphics{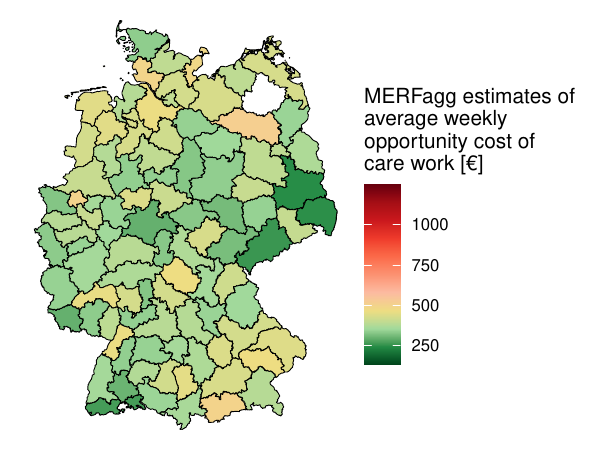}
	\caption{Spatial representation of area-level mean estimates from \textit{MERFagg} (\ref{MERFagg}) for mean weekly opportunity cost of care work [\EUR].}
	\label{fig:mapImportance}
\end{figure}

The results from the application of \textit{MERFagg} are reported in Figure \ref{fig:mapImportance}. We primarily focus on the technical details of the estimates produced by our proposed approach, postponing the contextual discussion of results to the end of this section. Overall, we observe a predominance of covariates such as age, gender, household size, and whether the person is employed in the public sector (cf. Table 3 in the online supplementary materials). We incorporate aggregated auxiliary information from $4$ up to $12$ covariates for the areas using optimal calibration-weights $\hat{w}_{ij}$.

\begin{figure}[tb]
	\centering
	\captionsetup{justification=centering,margin=1.5cm}
	\includegraphics[width=1\linewidth]{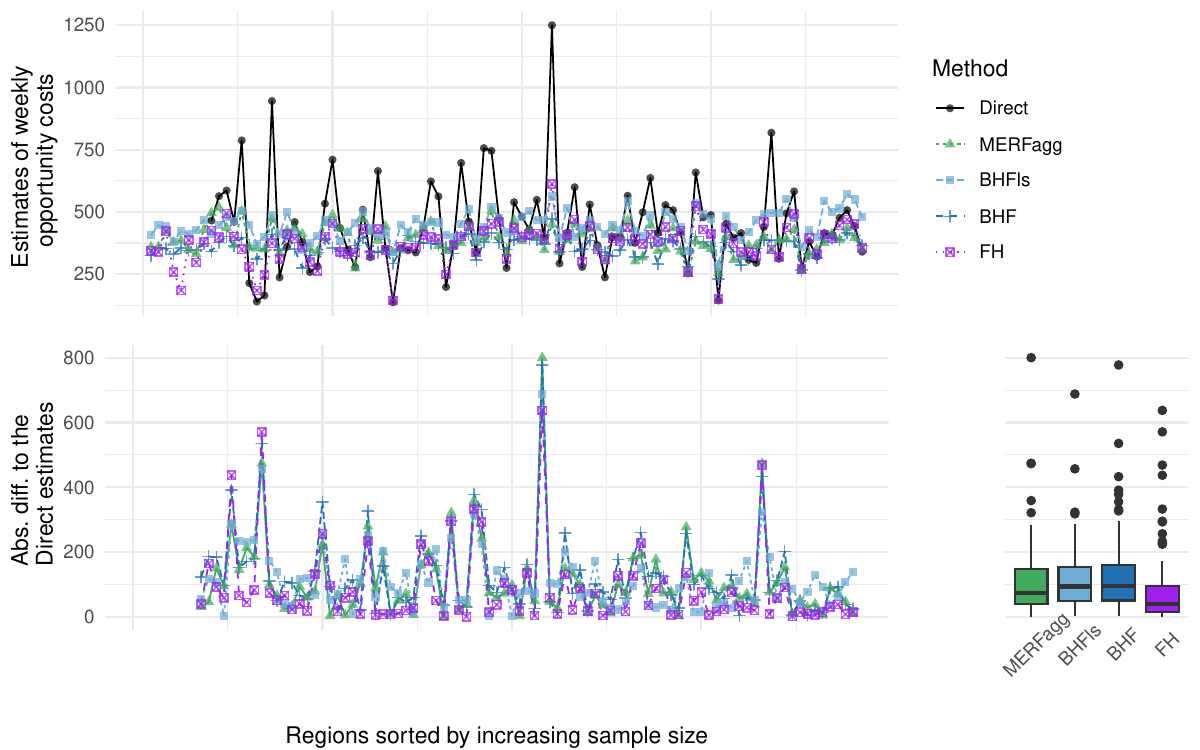}
	\caption{Comparison of area-level mean estimates for weekly opportunity cost of care work [\euro] across regions. The top panel shows the corresponding point estimates, while the bottom panel shows the absolute differences from the \textit{direct} estimates using a line plot and a boxplot.}
	\label{fig:Pointcompare}
\end{figure}

A comparison between the maps of \textit{direct} estimates in Figure \ref{fig:Overviewdirect} and estimates based on \textit{MERFagg} in Figure \ref{fig:mapImportance} indicates that \textit{MERFagg} results are more stable. Figure \ref{fig:Pointcompare} sorts the regions by increasing survey sample sizes, allowing for a more detailed discussion of the peculiarities in the point estimates for area-level means of weekly opportunity cost across $95$ regions. Estimates from the \textit{BHF} and \textit{BHFls} method are produced from the \textsf{R}-package \textsf{saeTrafo} \citep{saeTrafoPackage}, while the \textit{FH} estimates were obtained using the \textsf{R}-package \textsf{emdi} \citep{Harmening2023}. The MSEs for the \textit{BHF} and \textit{FH} estimates were calculated using analytical approximations \citep{Prasad_Rao1990}, whereas the MSE for the \textit{BHFls} was estimated by a parametric bootstrap with $B=200$ \citep{Wuerz_2022}. All four model-based estimators (\textit{MERFagg}, \textit{BHFls}, \textit{BHF}, and \textit{FH}) show a similar pattern in their deviation from the \textit{direct} estimates in Figure \ref{fig:Pointcompare}. These deviations decrease as sample size increase, and all four methods effectively track the high and low levels of the \textit{direct} estimates.

\begin{figure}[tb]
	\centering
	\captionsetup{justification=centering,margin=1.5cm}
	\includegraphics[width=1\linewidth]{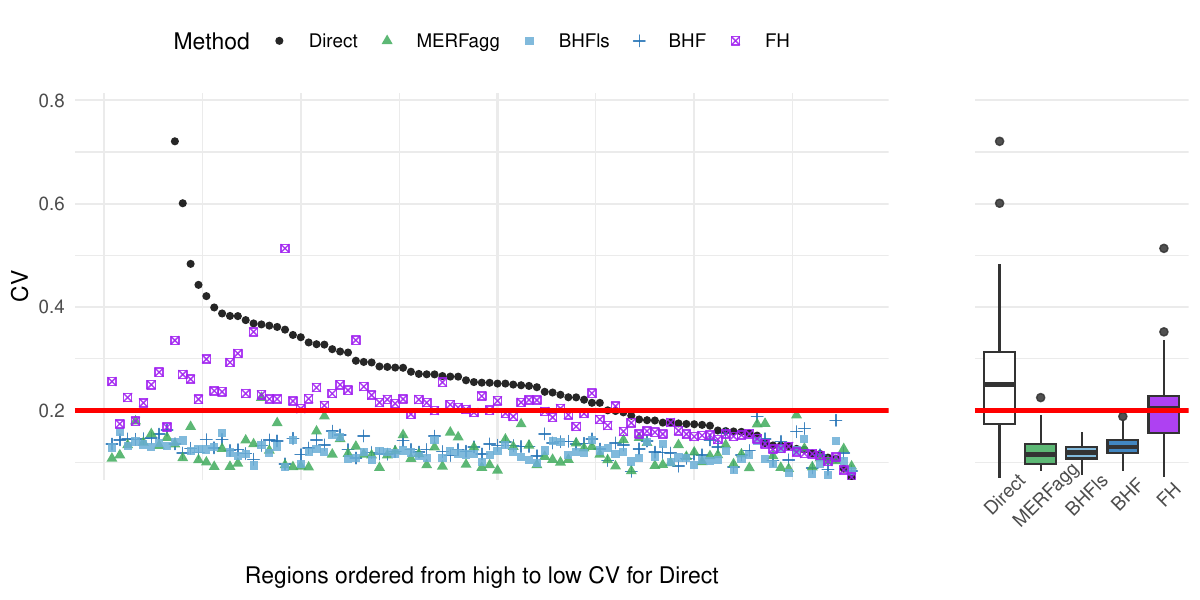}
	\caption{Comparison of area-specific CVs across regions. The red line marks the $20\%$-criterion for defining reliable estimates by \citet{eurostatCV} .}
	\label{fig:CVcompare}
\end{figure}

As previously discussed, \textit{direct} estimates suffer from relatively low accuracy, as indicated by their respective CVs. Figure \ref{fig:CVcompare} juxtaposes the CVs of \textit{direct} estimates with those of the \textit{BHFls}, \textit{BHF}, \textit{FH}, and our proposed \textit{MERFagg} method to contextualize the performance of the point estimates shown in Figure \ref{fig:Pointcompare}.
We observe that, on average, the CVs for all model-based estimators are smaller compared to those for \textit{direct} estimates. This improvement is particularly pronounced for regions with small sample sizes, where model-based estimators significantly reduce uncertainty. According to the boxplots in Figure \ref{fig:CVcompare}, \textit{MERFagg} demonstrates the lowest median CV ($11.46\%$), followed by \textit{BHFls} ($11.89\%$) and \textit{BHF} ($12.94\%$), though the differences are marginal. The \textit{MERFagg} estimates show a clear improvement over the \textit{direct} estimates: only $1$ region from $95$ fails to meet the required threshold of $20\%$. As expected, model-based estimates improve the accuracy, especially in regions with low sample sizes. Conversely, \textit{direct} estimates are relatively accurate for regions with high sample sizes. Please note that we further assess the application results by investigating the performance of the model-based estimators through a design-based simulation, as detailed in Section 4.2 of the online supplementary materials.

To further evaluate the plausibility of the \textit{MERFagg} results, we conduct two sensitivity checks. First, we compare the estimated random effects of the \textit{MERFagg} to those of the \textit{BHF}. Second, we assess two alternative calibration methods discussed in Section \ref{sec:2.3} and compare their performance to the proposed best-practice calibration strategy for the \textit{MERFagg}.

Figure 3 in the online supplementary materials shows the estimated random effects for both the \textit{MERFagg} and the \textit{BHF} across regions. The random effects of the \textit{MERFagg} closely track those of the \textit{BHF}, exhibiting similar patterns. However, the \textit{BHF} random effects display slightly greater variability, as indicated by higher variance estimates ($\hat{\sigma}_{u,MERF}=45.33$ vs. $\hat{\sigma}_{u,BHF}=50.62$). This difference reflects the \textit{MERFagg}'s ability to capture complex relationships between covariates, unlike the linear specification employed by the \textit{BHF}. This capability also explains the differences in area-level mean estimates between the \textit{MERFagg} and \textit{BHF}, as shown in Figure \ref{fig:Pointcompare}. Next, we compare the stability of the calibration weights derived from our framework, as discussed in Section \ref{sec:2.3}, with two alternative calibration methods. The first alternative uses ridge calibration weights \citep{rao1997ridge, bocci2008another, montanari2009multiple}, which are incorporated into the \textit{MERFagg} in Equation (\ref{MERFagg}) instead of the empirical likelihood-based weights proposed in our framework. The resulting MERF estimates are referred to as \textit{MERFagg\_ridge}, with calibration weights generated using the \textsf{R} package \textsf{icarus} \citep{icarus}. The second method computes calibration weights at the federal-state level, as the 95 regions are nested within the 16 federal states in Germany. These estimates are denoted as \textit{MERFagg\_fs}. Figure 4 in the online supplementary materials compares area-level mean estimates of the three calibration methods and their absolute differences from the \textit{direct} estimates. \textit{MERFagg} and \textit{MERFagg\_ridge} produce nearly identical results, because of the similarity of their weights. In contrast, \textit{MERFagg\_fs} exhibits slight deviations from the other two methods. Some covariates at the regional level are not homogeneous with the corresponding federal-state levels in which the regions are nested. We believe that these inhomogeneities may introduce the departures at the regional level in the current application.

Overall, regions across Germany report comparable levels of average individual weekly opportunity cost of care work. However, a detailed inspection of Figure \ref{fig:mapImportance} reveals a small cluster of lower values in the North-East of Germany. Explaining such patterns causally is challenging and may not yield effective conclusions. While wages and individual opportunity costs are directly related, the time spent for care work negatively affects opportunity cost. Thus, it is not observable whether these differences stem from variations in average income, increased time-allocation for care work, or a combination of both factors. Nevertheless, this approach enables us to identify and map the value of unpaid care work at a regional level in Germany.

\FloatBarrier
\section{Conclusion}\label{sec:6}
In this paper, we provide a coherent framework enabling the use of RFs for SAE under limited population-level auxiliary data. Our approach meets modern requirements of SAE, including the robustness against model-failure and aspects of data-driven model-selection within the existing methodological framework of SAE. 
The introduced semi-parametric unit-level mixed model treats traditional SAE methods as special cases. We address the challenging task of incorporating aggregated auxiliary information for MERFs and propose the use of calibration weights based on a profile empirical likelihood optimization problem. We deal with potential issues of numerical instabilities of the empirical likelihood approach and propose a best practice strategy for the application of our proposed estimator \textit{MERFagg} for SAE. The proposed point estimator for area-level means is complemented by a non-parametric MSE-bootstrap scheme. We evaluate the performance of point and MSE estimates compared to traditional SAE methods by a model-based simulation that reflects properties of real data (e.g., skewness). From these results, we conclude that our approach outperforms traditional methods in the existence of non-linear interactions between covariates and demonstrates robustness against distributional violations of normality for the random effects and for the unit-level error terms. The simulation results further demonstrate that incorporating area-specific random effects in MERFs enhances efficiency compared to a purely synthetic RF-based estimator. Moreover, we observe that the inclusion of aggregated auxiliary information through calibration weights based on empirical likelihood works reliably. Regarding the performance of our MSE-bootstrap scheme, we observe moderate levels of overestimation and report authentic tracking behaviour between estimated and empirical MSEs. We focus on a distinctive SAE example, where we study the average individual opportunity cost of care work for regions in Germany. Overall, we provide an illustrative example on how to use our data-driven best practice strategy on MERFs in the context of limited population-level auxiliary data. We identify a small cluster of lower levels of average individual opportunity cost of care work in the North-Eastern part of Germany.

From an empirical perspective, we face limitations that directly motivate further research. Firstly, we only calculate the opportunity cost of the working population and neglect care work done by people who already left the labour market due to care work issues. Despite its long tradition in economics, the basic concept of opportunity cost (treating the shadow value of care work equivalently to hourly wage from labour) faces drawbacks. Different models from a health and labour economic perspective (e.g., \cite{OlivaMoreno2019}) can be integrated into our approach. Nevertheless, given the data and our initial aim to provide a general methodology for regional mapping of care work specific differences, we consider the hourly wage as a first reasonable approximation to the unobservable ``real'' shadow price.

We motivate two major dimensions for further research, including theoretical considerations and aspects of generalizations. From a theoretical perspective, further research is needed to investigate the construction of a partial-analytical MSE for area-level means or the construction of an asymptotic MSE estimator. This also requires knowledge on the theoretical background for predictions from RFs \citep{Sexton_Laake2009, wager_etal2014, Wager_Athey2018, Athey_etal2019, Zhang2019}, existing research mainly aims to quantify the uncertainty of unit-level predictions. From a survey perspective, \citet{Dagdougetal2020} recently analyse theoretical properties of RFs in the context of complex survey data. The extension of these results for partly-analytical uncertainty measures in the context of dependent data structures and towards area-level indicators is non trivial and a conducive topic for theoretical SAE. Additionally, from a theoretical perspective, further exploration of calibration weights and the integration of survey weights into the MERF framework under aggregated auxiliary data is essential. Alternative methods to empirical likelihood-based calibration, such as ridge calibration \citep{bocci2008another} and box constraints \citep{theberge2000calibration}, could improve stability in situations with high variability in the weights. Model-calibrated estimators, as proposed by \cite{wu2001model} and \cite{park2009mixed}, incorporate weights that account for both the model structure and the complex sample design simultaneously. Although constructing weights that account for the model structure, the design, and other constraints aligns with \cite{deville1992calibration}, their integration into the MERF algorithm remains an open challenge.

Our approach shares the empirical likelihood-calibration-argument with \citet{Li_etal2019}, however, saves on the computationally intensive procedure of a smearing step \citep{Duan1983} without drawbacks on the predictive performance, because no transformations and corresponding bias exists. Nevertheless, we maintain that pairing our approach with a smearing argument allows for a more general methodology and subsequently for the estimation of indicators such as quantiles \citep{Chambers_Dunstan1986}. Apart from generalizations to quantiles, the approach of this paper is generalizable to model (complex) spatial correlations. Additionally, a generalization towards binary or count data is possible and left to further research. The semi-parametric composite formulation of Model (\ref{mod1}) allows for $f$ to adapt any functional form regarding the estimation of the conditional mean of $y_{ij}$ given $\mathbf{x}_{ij}$ and technically transfers to other machine learning methods, such as gradient-boosted trees or support vector machines.


\section*{Acknowledgements}
W\"urz gratefully acknowledges support by a scholarship of Studienstiftung des deutschen Volkes. Further, we would like to thank the Research Data Centre of the Statistical Offices of the federal states for providing 2022 census aggregates in Germany. The authors are grateful for the computation time provided by the HPC service of the Freie Universit\"at Berlin. Finally, the authors are indebted to the Joint Editor, Associate Editor and two referees for comments that significantly improved the paper.

\section*{Data availability statement}
The Socio-Economic Panel and the German Census data used in the application are not publicly available due to data disclosure agreements. The \textsf{R}-code for running the MERF models is included in the \textsf{SAEforest} package \citep{SAEforestPackage}. The \textsf{R}-scripts for reproducing the results in Section \ref{sec:4} are available upon request from the authors.

\bibliographystyle{apacite}			
\bibliography{./biblography}

\listoffigures

\end{document}


\maketitle
	\onehalfspacing
	\normalsize
	\setlength{\parindent}{0em}

\section{List of Abbreviations}
\begin{table}[h]
  \caption{List of Abbreviations.}
  \label{tab:abbreviations}
  	\footnotesize
 \centering
\begin{tabular}{@{\extracolsep{5pt}} lll} \\
[-1.8ex]\hline \hline \\[-1.8ex]
\textbf{Abbreviation}  & \textbf{Definition}&\textbf{Reference}\\\hline \\[-1.8ex]
BHF                  & Battese-Harter-Fuller&\cite{Battese_etal1988}\\
BHFls                & BHF estimator with an adaptive log-shift transformation&\cite{Wuerz_2022}\\
CV                   & Coefficient of Variation& \\
FH                   & Fay-Herriot&\cite{Fay_Heriot1979} \\
MERF                 & Mixed Effects Random Forest&Equation (1) in the paper \\
MERFagg              & MERF under aggregated auxiliary information& Equation (3) in the paper\\
                     & (weights are obtained by the proposed framework)& \\
MERFagg\_ridge       & MERF under aggregated auxiliary information& Equation (3) in the paper\\
                     & (weights are obtained by ridge calibration)& \\
MERFagg\_fs          & MERF under aggregated auxiliary information & Equation (3) in the paper\\
                     & (weights are obtained by calibration on federal-state level)& \\
MERFind              & MERF assuming access to unit-level auxiliary data& Equation (2) in the paper\\
MSE                  & Mean squared error& \\
RMSE                 & Empirical root mean squared error& Equation (6) in the paper\\
RB                   & Relative bias& Equation (7) in the paper\\
RB-RMSE              & Relative bias of RMSE& Equation (8) in the paper\\
RF                   & Random Forest& \cite{Breiman2001}\\
RFagg                & RF under aggregated auxiliary information& Equation (5) in the paper\\
SAE                  & Small Area Estimation& \\
TNER2                & Bias-corrected transformed nested error regression estimator& \cite{Li_etal2019}\\\hline \\[-1.8ex]
\end{tabular}
\end{table}

\section{Additional Information: Limitation of Empirical Likelihood and a Best Practice Advice for SAE (Section 2.3)}
The existence of a solution to the maximization problem for the calibration weights $\hat{w}_{ij}$ is not necessarily guaranteed for applications in SAE. A necessary and sufficient condition ensuring the existence of a solution for $\hat{\mathbf{\lambda}}_i$ is that the convex hull of the constraint matrix $\mathbf{x}_{ij} - \bar{\mathbf{x}}_{\text{pop},i}$ contains the origin as an interior point. Especially for small sample sizes $n_i$ this condition requires scrutiny \citep{Emerson_Owen2009}. If the sample means of $\mathbf{x}_{ij}$ for area $i$ strongly differ from $\bar{\mathbf{x}}_{\text{pop},i}$, for instance, due to a strong imbalance of the unit-level sample values $\mathbf{x}_{ij}$ around the area-specific mean from population data $\bar{\mathbf{x}}_{\text{pop},i}$, no optimal solution for $\hat{\mathbf{\lambda}}_i$ and subsequently $\hat{w}_{ij}$ can be obtained. The dimensionality of existing covariates $p$ relative to the sample size $n_i$ exacerbates the problem. As a result, the constraints in matrix $\mathbf{x}_{ij} - \bar{\mathbf{x}}_{\text{pop},i}$ are infeasible for finding a global optimum in Equation (4) in the paper. Concrete empirical examples are different largely unbalanced categorical covariates in $\mathbf{x}_{ij}$, leading to column-wise multicollinearity in the $n_i \times p$ matrix of constraints $\mathbf{x}_{ij} - \bar{\mathbf{x}}_{\text{pop},i}$.

Overcoming mentioned technical requirements, \citet{Li_etal2019} propose the use of the adjusted empirical likelihood approach by \citet{chen2008adjusted}, which forces the existence of a solution to Equation (4) in the paper. Essentially, the introduced adjustment is an additional pseudo-observation within each domain $i$, increasing area-specific sample sizes to $n_{i+1}$. This pseudo-observation is jointly calculated from respective area-specific survey and census means of the auxiliary covariates \citep{chen2008adjusted}. Although the added adjustment-observation reduces risks of numerical instabilities, it simultaneously imposes difficulties from an applied perspective of SAE. \citet{Emerson_Owen2009} scrutinize the application of adjusted empirical likelihood in the context of multivariate population means, maintaining that the added pseudo-observation distorts the true likelihood configuration even for moderate dimensions of $p$ in cases of low area-specific sample sizes $n_i$. \citet{chen2008adjusted} note, that the problem is mitigated if the semi-parametric model is correctly specified and if the initial estimates for $\bar{\mathbf{x}}_{{\text{smp}},i}$ are not too far away from the true population mean. Nevertheless, we observe that the influence of the bound-correction of \citet{chen2008adjusted} used by \citet{Li_etal2019} has drawbacks, which we will discuss in the model-based simulation in Section 4 in the paper.

Dealing with empirical examples characterized by low domain-specific sample sizes, we abstain from the approaches of adding synthetic pseudo-observations to each domain. We maintain that in the context of non-linear semi-parametric approaches (such as RFs) there is a risk of including implausible unit-level predictions from $f$ based on the pseudo-covariates, i.e. $\hat{y}_{\text{pseudo},i}$. In this sense, pseudo-observations manipulate the estimation of area-level means under limited population-level auxiliary information in two ways: indirectly through their effect on the determination of all weights $\hat{w}_{ij}$ and directly through the predicted pseudo-value that is added to the survey sample.

We postulate a stepwise approach to ensure a solution to Equation (4) in the paper for each area $i$ under a reduced risk of distortions driven by improper pseudo-values through optimization bound-corrections. This approach can be interpreted as a best-practice strategy on the incorporation of maximal auxiliary covariate information through calibration weights in Equation (4) in the paper for the estimation of area-level means with MERFs. 
In detail, we first check for each area $i$ whether perfect column-wise-dependence in the $p \times n_i$ matrix of constraints $(\mathbf{x}_{ij} - \bar{\mathbf{x}}_{\text{pop},i})_{j = 1, ..., n_i}$ exists. If so, we remove perfectly collinear columns and rerun the optimization. Subsequently, we proceed along two dimensions: a) increasing the sample size of $i$-th area and b) decreasing the number of auxiliary covariates $p$ to calculate $\hat{w}_{ij}$ for area $i$. For a) we advise to sample a moderate number of observations (e.g., $10$) randomly with replacement from an area which is ``closest'' to area $i$. We refer to areas as ``closest'', if they have the smallest Euclidean distance for the aggregated auxiliary information $\bar{\mathbf{x}}_{\text{pop},i}$. This additionally allows to handle out-of-sample areas. For b) we propose a backward selection of covariate information based on the variable importance. Variable importance are RF-specific metrics that enable the ranking of covariates reflecting their influence on the predictive model. As we are primarily concerned about the order of influence of covariates, we rank based on the mean decrease in impurity importance, which measures the total decrease in node-specific variance of the response variable from splitting, averaged over all trees \citep{Biau_Scornet2016}. Overall, our strategy to handle potential failure in the solutions for weights and out-of-sample domains is summarized in the following algorithmic strategy:

\par\noindent\rule{\textwidth}{0.4pt}
\begin{enumerate}
	\item Use MERF to obtain estimates $\hat{f}$, $\hat{u}_i$, $\hat{\sigma}_u^2$, and $\hat{\sigma}_e^2$ from available unit-level survey data and estimate the indicator $\hat{\mu}_i^{\text{MERFagg}}$ (3) including weights $\hat{w}_{ij}$ following Equation (4) in the paper.
	\item If the calculation of weights fails due to infeasibility of constraints in the optimization problem for area $i$:
	\begin{enumerate}
	\item Check the feasibility of constraints used in the optimization and remove perfectly co-linear columns in $(\mathbf{x}_{ij} - \bar{\mathbf{x}}_{\text{pop},i})_{j = 1, ..., n_i}$. Retry the optimization in Equation (4) in the paper.
	\item If the calculation of weights fails again, optionally enhance the domain-specific sample size of area $i$ by sampling randomly with replacement from the most ``similar'' domain according to the minimal row-wise Euclidean distance between area-specific aggregated auxiliary  information $\bar{\mathbf{x}}_{\text{pop},i}$. Retry the calculation of weights $\hat{w}_{ij}$.
	\item If it fails again, reduce the number of covariates used for the calculation of weights for area $i$. Starting with the least influential covariate based on variable importance from $\hat{f}$, reduce the number of covariates in each step and retry the calculation of weights after each step.
	\item If the calculation of weights was not possible in step (c), set $\hat{w}_{ij}$ to $1/n_i$. These weights are non-informative for incorporating auxiliary information, however, the model-based estimates $\hat{f}(\mathbf{x}_{ij}) + \hat{u}_i$ still comprise information from other in-sample areas.
	\end{enumerate}
	\item Calculate the indicator for the $i$-th area as proposed by Equation (3) in the paper.
\end{enumerate}
\par\noindent\rule{\textwidth}{0.4pt}

The general performance of the best-practice strategy is illustrated by the results of the model-based simulation in Section 4 in the paper. Additionally, the proposed best-practice strategy is applied in the application discussed in Section 5 in the paper.
\newpage
\section{Additional Information: Model-Based Simulation (Section 4)}
\begin{table}[!ht]
	\footnotesize
	\centering
	\caption{Mean and Median of RB and RMSE over areas for point estimates in four scenarios.}
	\begin{tabular}{@{\extracolsep{5pt}} lrrrrrrrrr}
		\\[-1.8ex]\hline
		\hline \\[-1.8ex]
		& &\multicolumn{2}{c}{\textit{Normal}} &\multicolumn{2}{c}{\textit{Pareto}}&\multicolumn{2}{c}{\textit{Interaction}}&\multicolumn{2}{c}{\textit{Logscale}} \\
		\hline \\[-1.8ex]
		& & Median & Mean & Median & Mean & Median & Mean & Median & Mean \\
		\hline \\[-1.8ex]
		\multicolumn{9}{l}{RB}\\
		\hline \\[-1.8ex]
		&MERFind & $0.0014$ & $0.0019$ & $0.0033$ & $0.0038$ & $0.0071$ & $0.0061$ & $0.0076$ & $0.0082$ \\
		&MERFagg & $0.0001$ & $0.0005$ & $0.0011$ & $0.0016$ & $0.0034$ & $0.0138$ & $0.0004$ & $0.0002$ \\
		&RFagg & $0.0084$&$0.0107$ & $0.0105$ & $0.0120$ & $0.0338$ & $0.0112$ & $0.0222$ & $0.0223$\\
		&BHFls & $0.0087$ &$0.0082$ & $0.0085$ & $0.0076$ & $0.0094$ & $0.0270$& $0.0083$ & $0.0094$\\
		&BHF&$0.0009$ & $0.0013$ & $0.0019$ & $0.0022$ & $0.0031$ & $0.0233$ & $-0.0188$ & $-0.0225$ \\		
		&TNER2 & $0.0002$ & $-0.0001$ & $-0.0003$ & $-0.0008$ & $0.0010$ & $0.0187$ & $-0.0014$ & $-0.0020$ \\
		&FH& $0.0045$ & $0.0044$ & $0.0012$ & $0.0013$ & $-0.0001$ & $0.0162$& $-0.0107$ & $-0.0112$ \\	
		\hline \\[-1.8ex]
		\multicolumn{9}{l}{RMSE}\\
		\hline \\[-1.8ex]
		& MERFind & $210.98$ & $216.52$ & $209.54$ & $211.55$ & $196.10$ & $210.16$ & $226.39$ & $230.30$ \\
		& MERFagg & $203.49$ & $234.35$ & $204.08$ & $233.28$ & $187.50$ & $214.12$ & $216.39$ & $236.63$ \\
		&RFagg&$506.44$ & $518.20$ & $444.57$ & $459.20$ & $433.54$ & $432.03$ & $485.71$ & $489.23$\\
		&BHFls& $209.22$ & $219.71$ & $178.76$& $185.17$ & $247.55$ & $277.27$ & $185.72$ & $197.14$\\
		& BHF & $196.66$ & $201.80$ & $194.99$ & $202.99$ & $244.99$ & $273.51$ & $278.02$ & $290.33$ \\
		& TNER2 & $422.73$ & $429.09$ & $443.33$ & $442.83$ & $246.50$ & $280.17$ & $316.78$ & $325.10$ \\
		& FH& $371.18$ & $375.37$ & $365.65$ & $367.71$ & $245.11$ & $278.74$ & $497.06$ & $504.72$\\
		\hline \\[-1.8ex]
	\end{tabular}
	\label{tab:MBpoint}
\end{table}

\begin{figure}[h]
	\centering
	\captionsetup{justification=centering,margin=1.5cm}
	\includegraphics[width=0.9\linewidth]{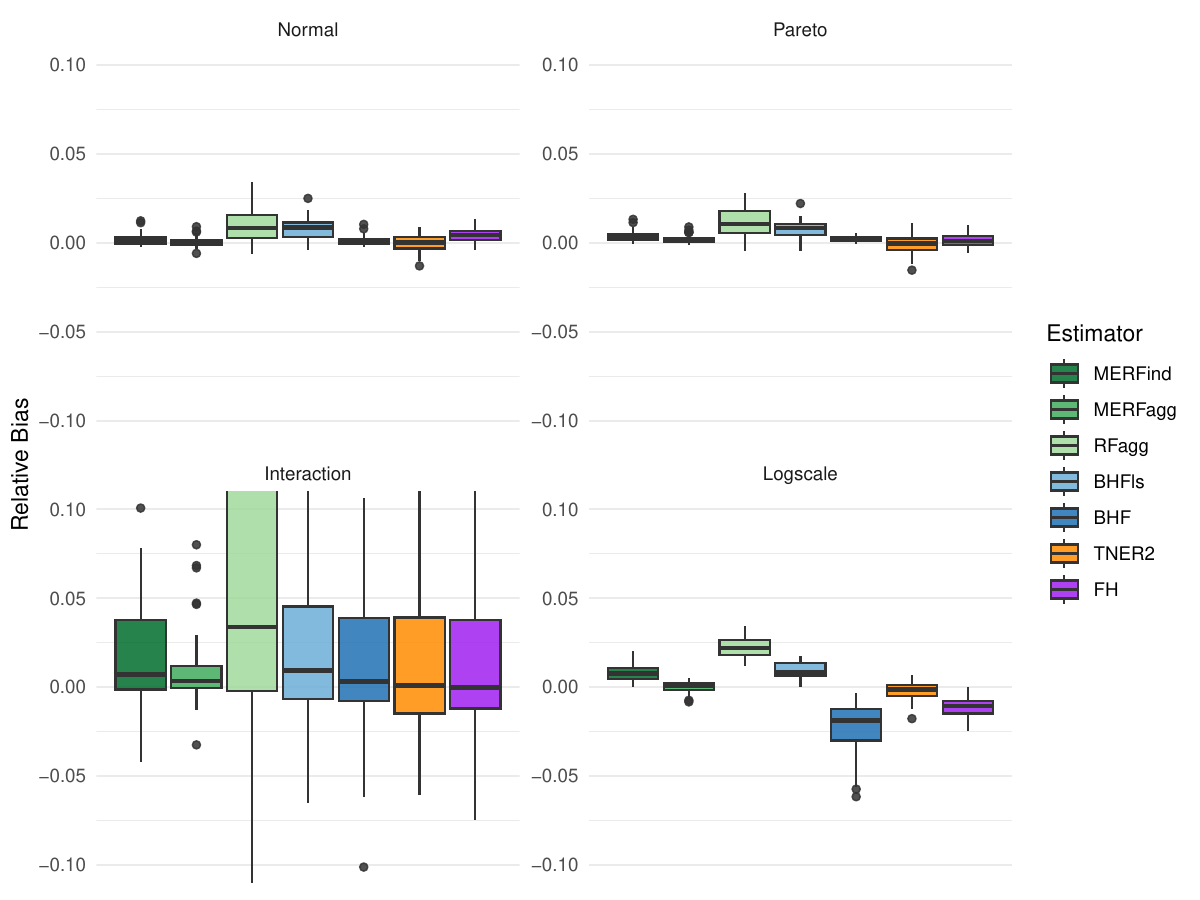}
	\caption{RB comparison of point estimates for area-level averages under four scenarios.}
	\label{fig:MBpoint_Bias}
\end{figure}

\begin{figure}[h]
	\centering
	\includegraphics[width=0.9\linewidth]{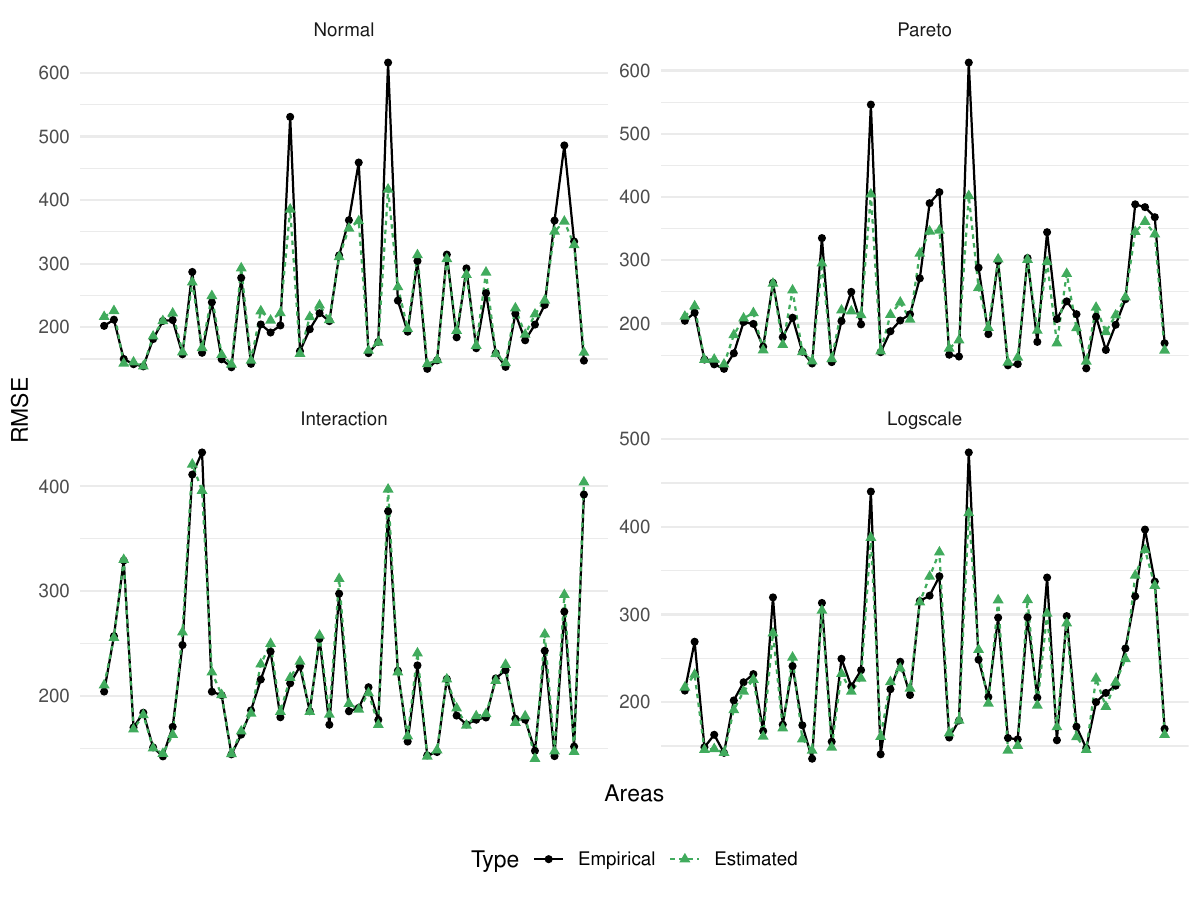}
	\caption{Estimated and empirical area-level RMSEs for four scenarios}
	\label{fig:trackMSE}
\end{figure}

\FloatBarrier
\newpage

\section{Additional Information: Application (Section 5)}
\subsection{Additional tables and plots}
\begin{table}[h!]\captionsetup{justification=centering,margin=1cm}
  \caption{Auxiliary variables on personal and socio-economic background and their variable importance based on the trained RF $\hat{f}$.}
  \label{tab:Covariates}
  	\footnotesize
 \centering
\begin{tabular}{@{\extracolsep{5pt}} lr} \\
[-1.8ex]\hline \hline \\[-1.8ex]
\textbf{Covariates}  & \textbf{Variable importance}\\\hline \\[-1.8ex]
 Age in years & $19541192.683$\\
 Sex & $5705960.532$ \\
 Number of persons living in household &  $5449178.802$\\
 Employment status: civil servants & $5284275.244$\\
 Migration background: direct & $3850417.870 $\\
 Grouped nationality: Asia & $2262376.935$\\
 Tenant or owner &$1882946.130$\\
 Position in Household: marriage-like & $1463114.089$\\
 Migration background: indirect & $1425117.242$ \\
 Position in Household: single parent & $1206034.041$ \\
 Employment status: employed without & $1120370.166$\\ \hspace{1cm}national insurance (e.g. mini-jobber) &\\
 Grouped nationality: European Union (excluding Germany) & $1075998.926$\\
 Grouped nationality: North America & $878752.597 $ \\
 Grouped nationality: remaining European countries & $468653.092$\\
 Position in Household: living alone & $568725.503$\\
 Grouped nationality: South America & $223675.151 $\\
 Grouped nationality: Africa & $146252.765 $\\
 Grouped nationality: Australia & $5996.311 $\\
\hline \\[-1.8ex]
\end{tabular}
\end{table}

\begin{figure}[!h]
	\centering
	\includegraphics[width = 0.9 \linewidth]{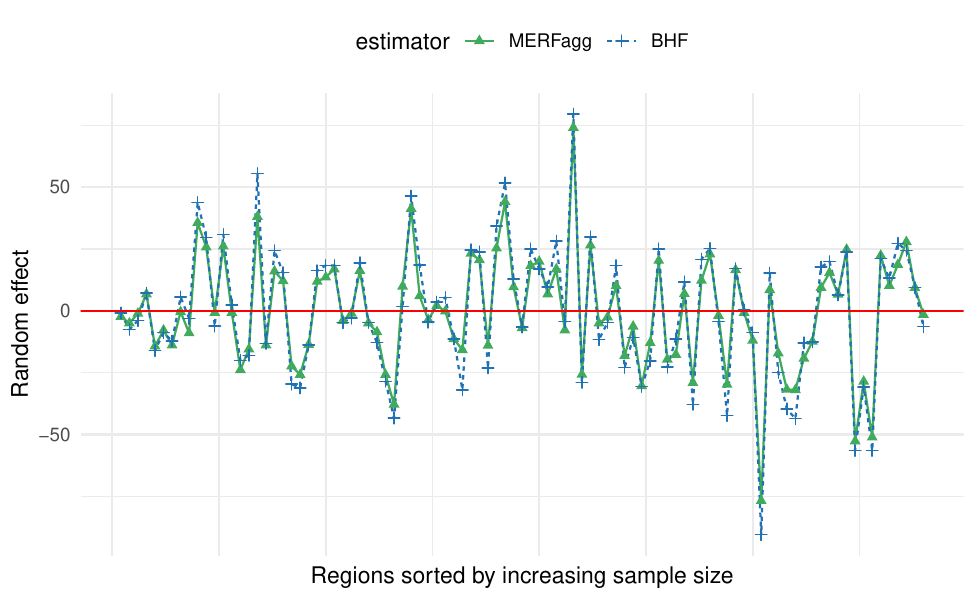}
	\caption{Estimated random effects of the \textit{MERFagg} and \textit{BHF} across regions.} \label{randomeff_linplot}
\end{figure}

\begin{figure}[h!]
	\centering
	\captionsetup{justification=centering,margin=1cm}
	\includegraphics[width=1.03\linewidth]{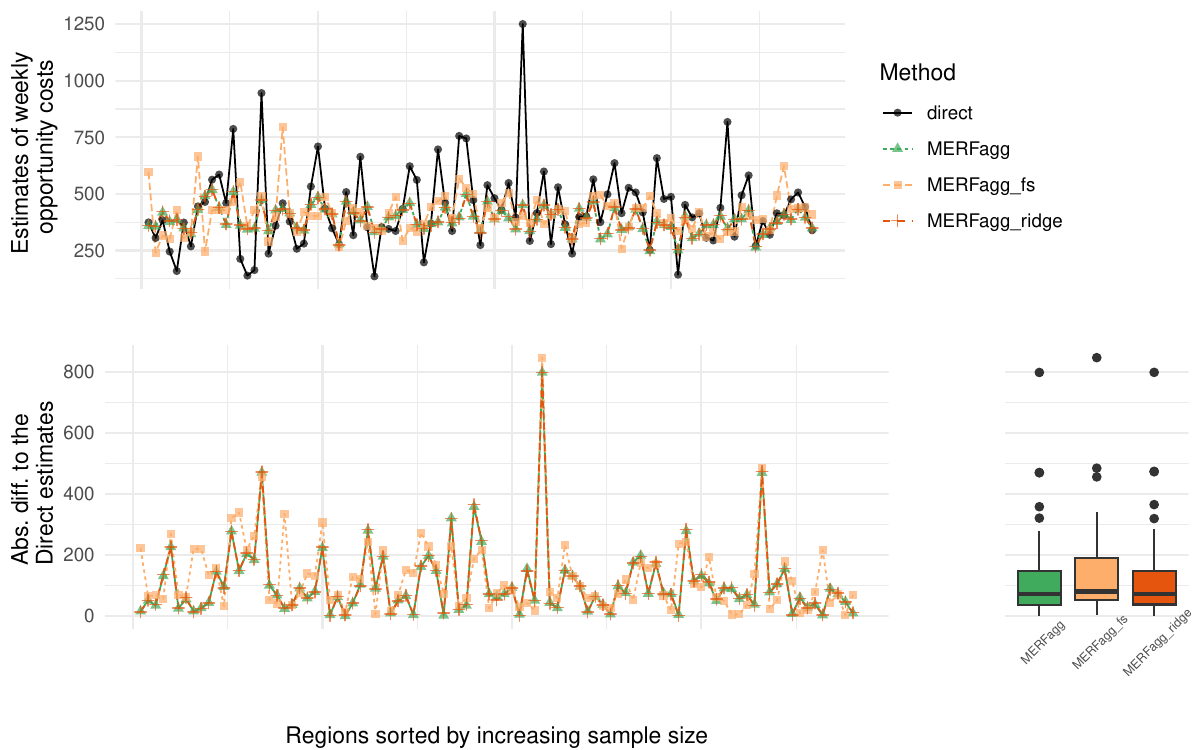}
	\caption{Comparison of area-level mean estimates for weekly opportunity cost of care work [\euro] across regions. The top panel shows the corresponding point estimates, while the bottom panel shows the absolute differences from the \textit{direct} estimates using a line plot and a boxplot.}
	\label{fig:trackMSE}
\end{figure}

\newpage
\FloatBarrier
\subsection{Design-based simulation}
In Section 4 of the paper, we conducted a model-based simulation to provide a controlled empirical assessment of the model-based methods for point and uncertainty estimation. In Section 5, the estimators requiring only aggregated population data were applied to the Socio-Economic Panel to estimate weekly opportunity costs at the regional level in Germany. To further evaluate the application results, we present a design-based simulation here to investigate the performance of the model-based estimators in a close-to-reality setting where the true data generation mechanism is unknown.

This study is also based on data from the 2021 Socio-Economic Panel \citep{SOEP_datasource}. We expand the survey data using individual survey weights to create an (artificial) population, from which 500 samples are drawn, reflecting the sample sizes of the 95 regions in the application (ranging from 3 to 145, with a mean of 31.37 and a median of 24). However, this approach is not without challenges. Specifically, the (artificial) population contains many duplicated individuals, which may reduce the variability of the data compared to the real-world application discussed in Section 5 of the paper.

We evaluate the performance of MERFs under limited auxiliary data access (\textit{MERFagg}, Equation (3) in the paper) against a synthetic estimator based on a classical random forest with limited auxiliary data access that does not account for area-specific random effects (\textit{RFagg}, Equation (5) in the paper) and the MERF assuming access to unit-level auxiliary data (\textit{MERFind}, Equation (2) in the paper). These estimators are compared to two alternatives that rely solely on aggregated population-level auxiliary data: (a) the \textit{BHF} estimator and (b) the area-level \textit{FH} estimator. The aggregated population-level auxiliary data are derived from the (artificial) population. Please note that we do not present results for the \textit{BHFls} estimator, as it requires not only census aggregates (means) but also region-specific covariance matrices of the covariates, which contain many zeros due to the duplicates in the (artificial) population.

\begin{figure}[t]
	\centering
	\captionsetup{justification=centering,margin=1cm}
	\includegraphics[width=1\linewidth]{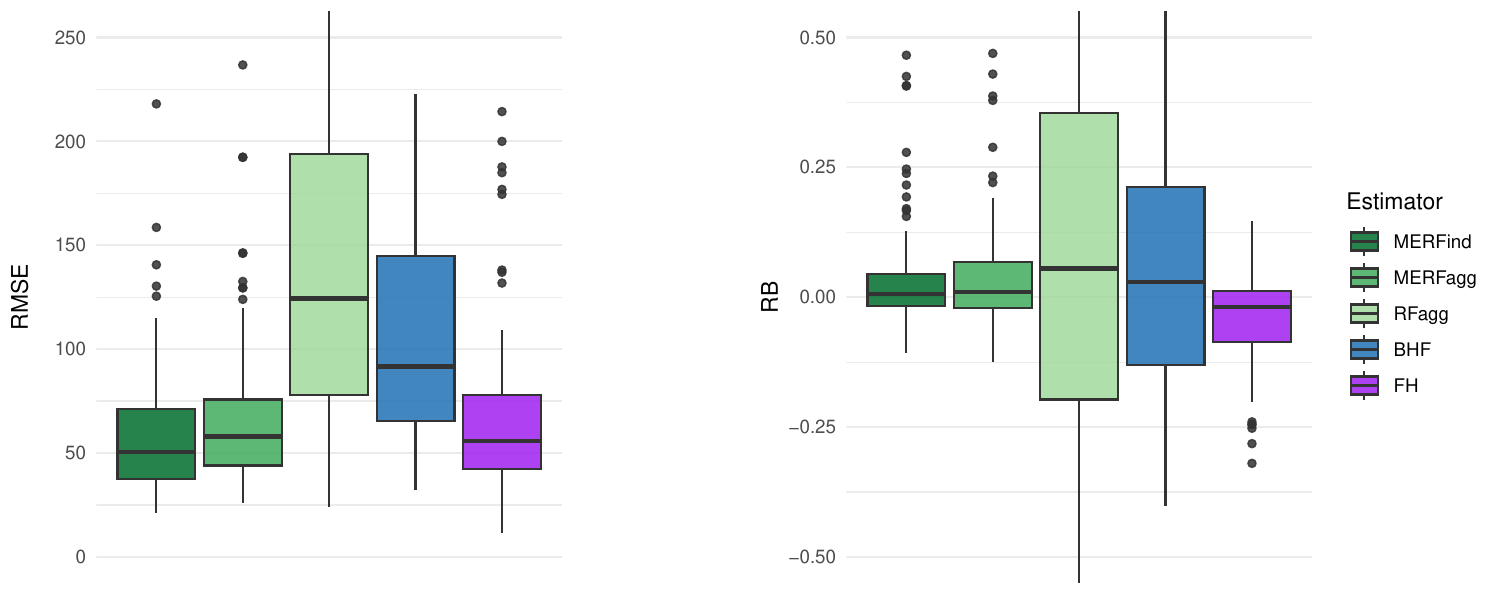}
	\caption{RB and RMSE of selected estimators in the design-based simulation across regions.}
	\label{fig:trackMSE}
\end{figure}
Figure \ref{fig:trackMSE} presents the RMSE (Equation (6) in the paper) and RB (Equation (7) in the paper) of the estimators in the design-based simulation across regions. The \textit{MERFind} and \textit{MERFagg} estimators perform comparably, exhibiting the lowest levels of uncertainty and bias. The proposed \textit{MERFagg} estimator outperforms both the \textit{BHF} and \textit{FH} estimators, which rely solely on aggregated population-level auxiliary data. This suggests that linear mixed model-based methods may struggle to capture the underlying predictive relationships between covariates in the application, whereas MERFs can effectively model non-linear relationships. Additionally, the \textit{MERFagg} estimator surpasses the \textit{RFagg} estimator, highlighting the advantage of incorporating random effects in this context. Finally, the similarity between \textit{MERFind} and \textit{MERFagg} suggests that the distinction between unit-level and aggregated auxiliary information has a minimal impact in the design-based simulation.

\bibliographystyle{apacite}			
\bibliography{./biblography}